\documentclass[fleqn,usenatbib]{mnras}
   
\usepackage{newtxtext,newtxmath}

\DeclareRobustCommand{\VAN}[3]{#2}
\let\VANthebibliography\thebibliography
\def\thebibliography{\DeclareRobustCommand{\VAN}[3]{##3}\VANthebibliography}

\usepackage{graphicx}	
\usepackage{amsmath}	
\usepackage{tgbonum}
\usepackage{comment}  
\usepackage{booktabs,caption}    
\usepackage{threeparttable} 
\usepackage{array} 
\usepackage{caption}
\usepackage{microtype}
\usepackage{chngcntr}
\counterwithin{table}{section}
\usepackage{rotating} 
\usepackage{tikz}
\usepackage{graphicx}
\usepackage{lipsum}
\usepackage{pdflscape}
\usepackage{float}
\usepackage{gensymb}
\usepackage{multirow}
\usepackage[T1]{fontenc}
\usepackage{ulem}
\usepackage{cancel}

\newcommand{\dotes}{\rlap{.}^\text{s}}

\newcommand{\dotarc}{\rlap{.}''}

\DeclareUnicodeCharacter{2212}{-}

\title[WISDOM XXVI.\ Cross-checking the SMBH of NGC~4751]{WISDOM Project -- XXVI.\ Cross-checking supermassive black hole mass estimates from ALMA CO gas kinematics and SINFONI stellar kinematics in the galaxy NGC~4751}

\author[P.\ Dominiak et al.]{Pandora Dominiak,$^{1}$\thanks{E-mail: pandora.dominiak@physics.ox.ac.uk}
Michele Cappellari,$^{1}$
Martin Bureau,$^{1}$\thanks{E-mail: martin.bureau@physics.ox.ac.uk}
Timothy A.\ Davis,$^{2}$
Marc Sarzi,$^{3}$
Ilaria Ruffa,$^{2}$
\newauthor{Satoru Iguchi,$^{4,5}$ Thomas G. Williams$^{1}$ and Hengyue Zhang$^{1}$}
\\
$^{1}$Sub-department of Astrophysics, Department of Physics, University of Oxford, Denys Wilkinson Building, Keble Road, Oxford, OX1~3RH, UK\\
$^{2}$Cardiff Hub for Astrophysics Research \&\ Technology, School of Physics \&\ Astronomy, Cardiff University, Queens Buildings, Cardiff, CF24~3AA, UK\\
$^{3}$Armagh Observatory and Planetarium, College Hill, Armagh BT61~9DG, UK\\
$^{4}$Department of Astronomical Science, SOKENDAI (The Graduate University of Advanced Studies), Mitaka, Tokyo 181-8588, Japan\\
$^{5}$National Astronomical Observatory of Japan, National Institutes of Natural Sciences, Mitaka, Tokyo 181-8588, Japan\\
}

\date{Accepted XXX. Received YYY; in original form ZZZ}

\pubyear{2025}

\begin{document}
\label{firstpage}
\pagerange{\pageref{firstpage}--\pageref{lastpage}}
\maketitle

\begin{abstract}
We present high angular resolution ($0.19$~arcsec or $\approx24$~pc) Atacama Large Millimeter/submillimeter Array observations of the $^{12}$CO(3–2) line emission of the galaxy NGC~4751. The data provide evidence for the presence of a central supermassive black hole (SMBH). Assuming a constant mass-to-light ratio ($M/L$), we infer a SMBH mass $M_\text{BH}=3.43^{+0.45}_{-0.44}[\text{stat},\,3\sigma]^{+0.22}_{-0.64}[\text{sys}]\times10^9$~M$_\odot$ and a F160W filter stellar $M/L_{F160W}=2.68\pm0.11[\text{stat},\,3\sigma]^{+0.10}_{-0.80}[\text{sys}]$~M$_\odot$/L$_{\odot,\text{F160W}}$, where the first uncertainties are statistical and the second systematic. Assuming a linearly spatially-varying $M/L$, we infer $M_\text{BH}=2.79^{+0.75}_{-0.57}[\text{stat},\,3\sigma]^{+0.75}_{-0.45}[\text{syst}]\times10^9$~M$_\odot$ and $\left(M/L_\text{F160W}\right)/\left(\text{M}_\odot/\text{L}_{\odot,\text{F160W}}\right)=3.07^{+0.27}_{-0.35}[\text{stat},\,3\sigma]^{+0.08}_{-1.14}[\text{sys}]-0.09^{+0.08}_{-0.06}[\text{stat},\,3\sigma]^{+0.08}_{-0.01}[\text{sys}]\,\left(R/\text{arcsec}\right)$, where $R$ is the galactocentric radius. We also present SMBH mass estimates using the Jeans Anisotropic Modelling (JAM) method and Very Large Telescope Spectrograph for INtegral Field Observations in the Near Infrared (SINFONI) stellar kinematics. Assuming a cylindrically-aligned velocity ellipsoid (JAM$_\text{cyl}$) we infer $M_\text{BH}=(2.52\pm 0.36)\times10^9$~M$_\odot$, while assuming a spherically-aligned velocity ellipsoid (JAM$_\text{sph}$) we infer $M_\text{BH}=(3.24\pm0.87)\times10^9$~M$_\odot$. The SMBH mass assuming a constant $M/L$ is statistically consistent with that of JAM$_\text{sph}$, whereas the mass assuming a linearly-varying $M/L$ is consistent with both JAM$_\text{cyl}$ and JAM$_\text{sph}$ (within the uncertainties). Our derived masses are larger than (and inconsistent with) one previous stellar dynamical measurement using the Schwarzschild orbit-superposition method and the same SINFONI kinematics.
\end{abstract}



\section{Introduction}
\label{section:Introduction}

Observations over the last three decades have demonstrated that nearly every massive galaxy hosts a supermassive black hole (SMBH) at its centre. These SMBHs dynamically influence only the most central regions of their host galaxies. Despite this, their properties are tightly correlated with those of their hosts \citep[e.g.][]{Magorrian1998, Kormendy2013, McConnell_2013}. The tightest of these correlations is that between SMBH mass ($M_\text{BH}$) and stellar velocity dispersion measured within one effective (i.e.\ half-light) radius ($\sigma_\text{e}$), commonly referred to as the $M_{\text{BH}}$ -- $\sigma_\text{e}$ relation \citep[e.g.][]{Gebhardt_2000, Gultekin_2009}. These correlations suggest that SMBHs co-evolve with their host galaxies, but the details of the self-regulating processes are still poorly understood.

The ability to study SMBH -- host galaxy correlations relies on accurate methods of SMBH mass determination. The most reliable estimates are obtained by probing matter within the gravitational spheres of influence (SoI) of the SMBHs, where the SMBH dominates the gravitational potential. These methods thus require high-spatial resolution  observations to discern the impacts of the SMBHs on the matter in the innermost regions of their host galaxies. While a variety of kinematic tracers exist to probe the SMBH SoI, such as stars, ionised gas and megamaser discs (referred to as masers throughout), different methods tend to work best in different types of targets. Stellar kinematic methods have mainly been used in early-type galaxies (ETGs), whereas ionised gas is typically used in late-type galaxies (LTGs), and masers are predominantly present in rather low-mass Seyfert~2 and low-ionisation nuclear emission region (LINER) galaxies. Additionally, each method suffers from distinct systematic weaknesses that might bias the derived SMBH masses \citep{Bosch_2016}.

The strength of stellar dynamical modelling is that it offers a ubiquitous probe of supermassive black hole (SMBH) masses, as all galaxies host stellar populations. However, the construction of robust dynamical models necessitates sophisticated modelling techniques and generally requires high signal-to-noise ratio (S/N) stellar kinematic data to measure the full shape of the LOSVD. Acquiring such data often demands large aperture telescopes or long integration times, particularly for low surface brightness galaxies such as massive cored galaxies. Dust extinction can also significantly impact kinematic measurements, frequently motivating observations in the near-infrared regime \citep[e.g.][]{Thater2019}. While dynamical models have historically often adopted an axisymmetric potential \citep[e.g.][]{Gebhardt2003}, a non-negligible fraction ($\approx12\%$) of ETGs with $M_*>10^{10}$~$\text{M}_\odot$ are classified as weakly triaxial slow rotators \citep{Emsellem_2011}. Triaxial models \citep[e.g.][]{vdBosch_2008} may be formally required for these systems, although early investigations have indicated that the recovered SMBH masses are consistent with those derived from axisymmetric models within the quoted uncertainties \citep{vdBosch_2010}. Furthermore, stellar dynamical models employing Schwarzschild's orbit-superposition method must describe the galaxy potential to large radii, necessitating the inclusion of a dark matter halo component, which can potentially increase the overall systematic uncertainties of the SMBH mass measurements \citep{Gebhardt_Thomas_2009}.

Whilst modelling the ionised gas is conceptually straightforward, relying on the rotational velocities of a dynamically-cold thin disc, it presents a different set of difficulties. Firstly, the assumption of a dynamically-cold thin disc is seldom borne out by the data, as ionised gas is particularly susceptible to non-gravitational forces such as turbulence, shocks and radiation pressure. It is therefore no surprise that most galaxies, especially disc systems, do not fulfil the requirements for easy ionised gas modelling. It is a long-standing issue that ionised gas methods yield systematically lower SMBH mass estimates than stellar kinematic methods due to pressure support \citep{Kormendy2013}.

Similarly, despite maser observations generally offering the best resolution of the SMBH SoI, and thus being considered the gold-standard for SMBH mass determination, the method is not without its problems. Maser observations yield the velocities of only a few spots along the major axis of the disc, effectively yielding one-dimensional kinematics. The lack of two-dimensional kinematic information makes it difficult to assess the internal structure of the disc for warps and other non-circular motions \citep[e.g.][]{Greenhill_2003}, which themselves can affect the determined SMBH mass.

The fact that SMBH mass measurements vary so starkly between modelling methods has consequences on our understanding of SMBH -- host galaxy relations. Firstly, we do not know how much of the observed scatters in SMBH -- host galaxy relations is intrinsic or a consequence of the uncertainties of the SMBH masses. This is a particularly pressing issue, as the best-fitting slopes of the ETG and LTG relations differ significantly \citep{Lauer_2007,McConnell_2013}, but as discussed above so do the methods of SMBH mass determination in galaxies of different morphological types. Due to sparser measurements, there are also significant uncertainties in the slopes, scatters and forms of the relations towards the high-mass end \citep{McConnell_2013}. This issue is exacerbated by the fact that many of the aforementioned simplifying assumptions add to the SMBH mass uncertainties. Lastly, there is evidence of a divergence between the two most fundamental SMBH -- host galaxy relations: the $M_{\text{BH}}$ -- $\sigma_\text{e}$ and the $M_{\text{BH}}$ -- bulge luminosity ($L_\text{bul}$) relation, whereby the former predicts far fewer massive SMBHs ($M_\text{BH}>10^9$~M$_\odot$; \citealt{Walsh2013}, \citealt{Bosch_2016}). It is clear that one of these relations must at the very least be more fundamental than the other, and may even be wrong at the high-mass end.

To identify potential inconsistencies between different methods, uncover systematic SMBH mass biases and resolve the uncertainties in SMBH -- host galaxy relations, there is an acute need to cross-check SMBH masses obtained using different methods in the same targets. However, such cross-checks have proven to be incredibly challenging due to the limited number of targets suitable for multiple methods of mass determination. To date, the SMBH masses of only $11$ targets have been cross-checked between some of the stellar, ionised gas, reverberation mapping and maser methods (see \citealt{Liang_2023} for a summary).

In recent years, a new method of SMBH mass determination has emerged that utilises cold molecular gas as the kinematic tracer. Particularly, CO has come to dominate the field. Attempts at using other molecular gas tracers such as hot molecular hydrogen have proven difficult due to turbulence and excitation from accretion and jet heating \citep[e.g.][]{Scharwächter_2013}. CO is a good dynamical tracer as it can be detected in a wide variety of galaxies along the Hubble sequence (even those that are no longer star forming), it is generally dynamically cold and, in the case of low-$J$ rotational transitions, the observations (and the derived kinematics and dynamical masses) are unaffected by dust. It is however worth noting that subsequent modelling steps still require the use of an optical or near-infrared image to quantify the stellar contribution of each galaxy, and this can be affected by dust (thus potentially affecting the fraction of the total dynamical mass attributed to the stars and the SMBH). The molecular gas method has been used most often in typical ETGs (\citealt{Davis_2013b, Barth2016, Davis2017, WISDOM_I, WISDOM_III, Nagai2019, Boizelle2019, WISDOM_V, WISDOM_IV, Ruffa_2019, Davis_2020, Cohn_2021, Smith_2021, Boizelle_2021, Kabasares_2022, Ruffa_2023, Dominiak_2024, Zhang_2025}), but it has also been used in three LTGs (all barred spirals; \citealt{Onishi_2015, Nguyen_2020, Nguyen_2021}), a dwarf ETG \citep{Davis_2020} and a peculiar luminous infrared galaxy (LIRG) with central spiral arms \citep{Lelli_2022}.

The variety of targets in which this method can be utilised makes molecular gas SMBH mass measurements promising candidates for cross-checks. Thus far, molecular gas has enabled the following cross-checks in $7$ objects, increasing the number of cross-checked masses by two thirds: (i) `molecular gas vs.\ ionised gas' in NGC~4261 \citep{Ferrarese_1996, Boizelle_2021} and NGC~7052 \citep{vanderMarel_1998, Smith_2021} and (ii) `molecular gas vs.\ stars' in NGC~524 \citep{Krajnovic2009, WISDOM_IV}, NGC~1332 \citep{Rusli_2010, Barth2016}, NGC~4697 \citep{Schulze_2011, Davis2017}, NGC~6861 \citep{Rusli2013, Kabasares_2022} and the dwarf galaxy NGC~404 \citep{Nguyen_2017, Davis_2020}.

In this paper we present observations and kinematic modelling of the $^{12}$CO(3–2) line emission of the ETG NGC~4751, observed at high angular resolution with the Atacama Large Millimeter/submillimeter Array (ALMA). NGC~4751 already has a SMBH mass determined using Schwarzschild orbit-superposition modelling of its stellar kinematics \citep{Rusli2013}. In this paper we present an alternative stellar kinematic SMBH mass estimate using the same data as \citet{Rusli2013} but the Jeans Anisotropic Modelling (JAM) method \citep{Cappellari2008,Cappellari_2020}. Not only will these new measurements allow us to add NGC~4751 to the growing list of SMBH masses cross-checked between the molecular gas and stellar kinematic methods, but they will also allow us to compare SMBH masses determined using different stellar kinematic modelling methods.

The structure of this paper is as follows. In Section~\ref{section:NGC4751} we discuss the target and its relevant properties. In Section~\ref{section:ALMA Observations} we present the ALMA data and properties of the resulting CO data cube and continuum emission. We describe and present the results of the molecular gas dynamical modelling in Section~\ref{section:Dynamical Modelling}, whereas in Section~\ref{section:JAM models} we present the alternative SMBH mass measurements using stellar kinematics. We discuss these results within the context of other SMBH mass measurements in the literature in Section~\ref{section:Discussion}, and summarise and conclude in Section~\ref{section:Conclusion}. In Appendix~\ref{section:Appendix} we provide a comparison of the CO data with our best-fitting model with a constant stellar mass-to-light ratio $M/L$.


\section{NGC~4751}
\label{section:NGC4751}

NGC~4751 is an ETG located at $12^{\text{h}}52^{\text{m}}50\dotes79$, $-42\degree39\arcmin35\farcs7$ (J2000.0). In this paper we adopt a distance $D=26.9$~Mpc, that was calculated using the galaxy heliocentric velocity corrected for infall of the Local Group into the Virgo Cluster of galaxies and assuming a Hubble constant $H_0=72$~km~s$^{-1}$~Mpc$^{-1}$ \citep{Rusli2013}. We choose the same distance and cosmology as \citet{Rusli2013}, allowing us to perform a direct comparison of the SMBH mass measurements. At this distance, $1$~arcsec corresponds to $\approx130$~pc.  

NGC~4751 has a total absolute $B$-band magnitude of $−19.71$ and a total absolute $V$-band magnitude of $−20.75$ \citep{Rusli2013}. Based on Two Micron All Sky Survey $K_\text{s}$ filter images, it has a spheroid absolute magnitude of $−21.53\pm0.60$ and a total absolute magnitude of $−22.11\pm0.20$ \citep{Sahu_2019}. Based on these and a $3.6$~$\mu\text{m}$ stellar $M/L$ of $0.7$~M$_\odot/$L$_{\odot,3.6\mu\text{m}}$ \citep{Sahu_2019}, NGC~4751 has an estimated spheroid stellar mass $M_{\star
,\text{sph}}=3.09^{+2.53}_{-1.39}\times10^{10}$~M$_\odot$ and an estimated total galaxy stellar mass $M_{\star,\text{gal}}=5.25^{+1.67}_{-1.27}\times10^{10}$~M$_\odot$ \citep{Sahu_2019}. 

\begin{figure*}
    \centering
    \includegraphics[scale=0.51, trim={0.5cm 0 0 0}]{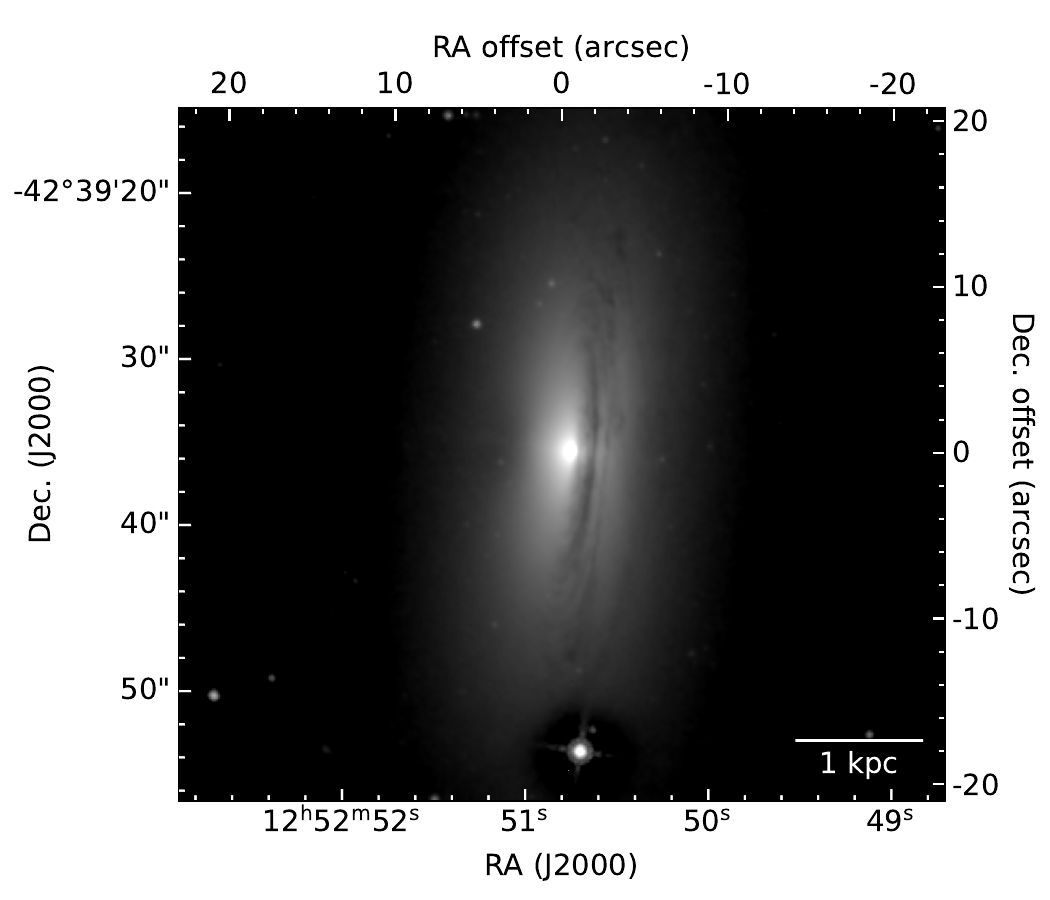}
    \includegraphics[scale=0.51, trim={0.2cm 0 0 0}]{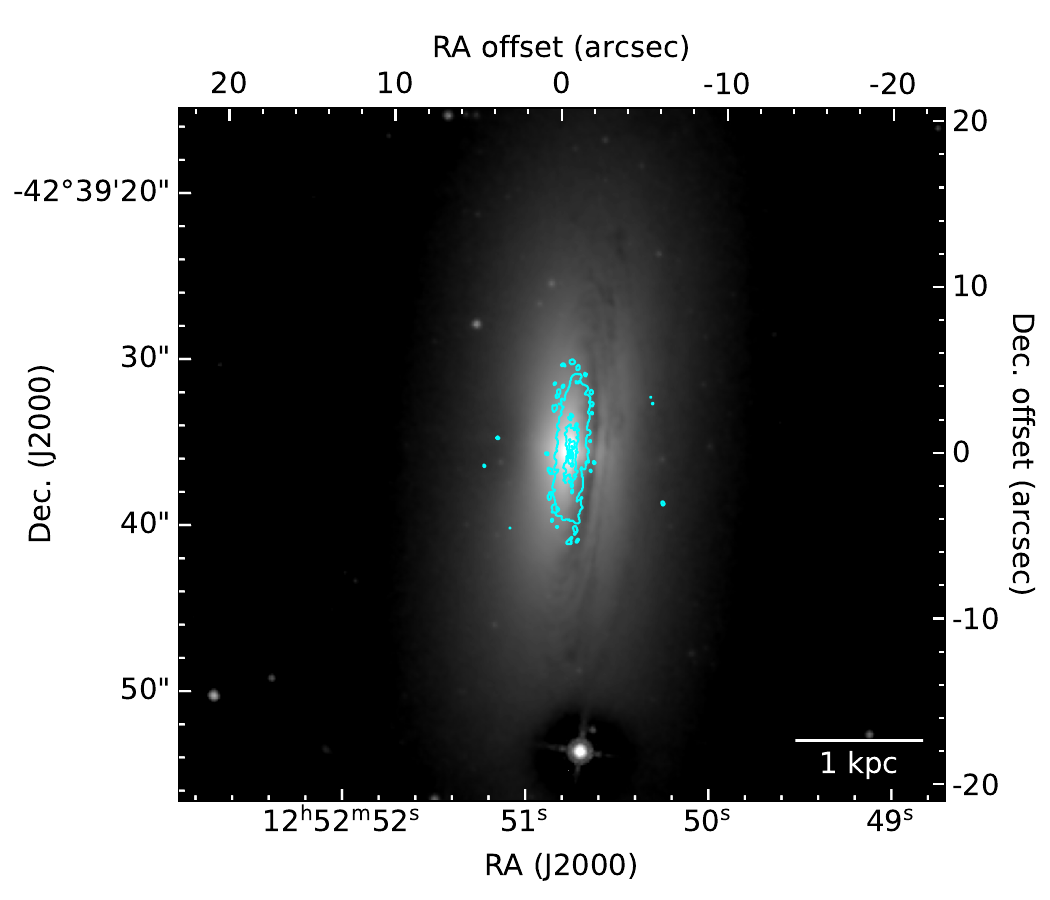}
    \caption{Unsharp-masked \textit{HST} Wide Field Camera 3 F160W filter ($H$-band) image of NGC~4751 (left), overlaid with the $^{12}\text{CO(3–2)}$ integrated-intensity contours (cyan) from our ALMA observations (right). A scale bar is shown in the bottom-right corner of each panel.}
    \label{fig:HST}
\end{figure*}

\textit{Hubble Space Telescope} (\textit{HST}) optical images reveal a prominent nearly edge-on dust disc, $\approx34$~arcsec in diameter along its major axis, with dust lanes to the west of an unobscured nucleus (see Fig.~\ref{fig:HST}). There is a small object $\approx18$~arcsec south of NGC~4751 that is most likely a foreground star (Fig.~\ref{fig:HST}). 

There is some disagreement as to whether NGC~4751 is an elliptical or a lenticular galaxy. Images of NGC~4751 suggest that it is comprised of two components, a relatively round central component with a steep radial surface brightness profile and an outer component that is flatter and has a shallower surface brightness profile. This is a defining characteristic of S0 galaxies, and if the central component is interpreted as a bulge and the outer component as a disc, the bulge-to-total luminosity ratio $B/T=0.55\pm0.05$ \citep{Kormendy2013}. Thus, canonically-speaking, NGC~4751 should be classified as an S0 galaxy. However, some studies suggest that both cluster \citep{KFCB} and field \citep{Huang_2013} ellipticals are naturally divided into objects with cored stellar profiles that formed through dry (gas-poor) mergers and cuspy stellar profiles that formed through wet (gas-rich) mergers. Thus, \citet{Kormendy2013} argued that NGC~4751 is really a highly-flattened extreme (E6) cuspy elliptical galaxy, whereby the central component is interpreted as resulting from a starburst event and the outer component is the consequence of the violent relaxation of pre-existing stars \citep{Mihos_1994, Hopkins_2009}.  

Integral-field spectroscopic data obtained using adaptive optics-assisted SINgle Faint Object Near-IR Investigation (SINFONI) observations on the Very Large Telescope were modelled using three-integral Schwarzschild models by \citet{Rusli2013} to infer a central SMBH mass $M_\text{BH}=(1.4\pm0.1)\times10^9$~M$_\odot$ and an $R$-band stellar $M/L_R=12.2^{+0.6}_{-0.7}$~M$_\odot/$~L$_{\odot,R}$ (based on a model with dark matter present) or $M/L_R=13.1^{+0.3}_{-0.4}$~M$_\odot/$~L$_{\odot,R}$ (based on a model without dark matter).

NGC~4751 has an effective radius $R_\text{e}=22.8$~arcsec, a central velocity dispersion $\sigma_0=357.6\pm17.7$~km~s$^{-1}$ \citep{Campbell_2014} and an effective stellar velocity dispersion $\sigma_\text{e}=355.4\pm13.6$~km~s$^{-1}$ \citep{Rusli2013}. Using the latter and the $M_{\text{BH}}$ -- $\sigma_\text{e}$ relation of \citet{Bosch_2016}, we expect a SMBH mass of $\approx4.5\times10^9$~M$_\odot$, about three times larger than that inferred by \citet{Rusli2013}. The radius of a SMBH's sphere of influence ($R_{\text{SoI}}$) quantifies the spatial extent over which the gravitational potential of the SMBH is dominant. One useful estimate of $R_{\text{SoI}}$ is $GM_{\text{BH}}/\sigma_\text{e}^2$, where $G$ is the gravitational constant. Using the SMBH mass of \citet{Rusli2013} yields $R_\text{SoI}\approx48$~pc ($\approx0.37$~arcsec), whereas the SMBH mass estimate from the $M_{\text{BH}}$ -- $\sigma_\text{e}$ relation yields $R_\text{SoI}\approx154$~pc ($\approx1.18$~arcsec).


\section{ALMA Observations}
\label{section:ALMA Observations}

The $^{12}\text{CO(3–2)}$ emission line of NGC~4751 was observed using the $12$-m ALMA array in band~7, as part of project 2016.1.01135.S (PI: Nagar). The data were collected with one track on 2017 May 17, for a total of $1361$~s on source. The baselines ranged from $15$~m to $1.1$~km, a maximum recoverable scale of $2.2$~arcsec ($\approx290$~pc) and a field of view of $16.7$~arcsec ($\approx2.2$~kpc). 

The observations had four spectral windows, each with a bandwidth of $2.0$~GHz ($\approx1735$~km~s$^{-1}$) subdivided into $128$ channels of $\approx16$~MHz ($\approx13.5$~km~s$^{-1}$). Two of the spectral windows were centred on both sides of the redshifted frequency of the $^{12}\text{CO(3–2)}$ line (rest frequency $\nu_{\text{rest}}=345.796$~GHz), with a small gap in frequency between them. The remaining two spectral windows were used to map the continuum. The data were calibrated using the \textsc{Common Astronomy Software Applications} (\textsc{casa}) package version 4.7.2 \citep{McMullin} and its ALMA pipeline. The following imaging steps used \textsc{casa} version 6.4.3.


\subsection{Line emission}
\label{section:line emission}

We remove the continuum emission by linearly fitting the channels of the continuum spectral windows and the line-free channels of the line spectral windows, and subtracting the best fit from the data in the \textit{uv}-plane using the \textsc{casa} task \texttt{uvcontsub}. Using the \textsc{casa} task \texttt{tclean}, we interactively cleaned the continuum-subtracted \textit{uv} data to a threshold equal to the root-mean-square (RMS) noise of the dirty channels in the cube (in regions without line emission). We imaged the data using the same \textsc{casa} task, Briggs weighting with a robust parameter of $0.5$ and a channel width of $30$~km~s$^{-1}$. We chose this channel width, roughly twice the native one, to improve the signal-to-noise ratio of the emission within the cube, in turn allowing us to adopt a higher angular and thus physical resolution, as required by our science goals. While the minimum SMBH mass detectable tends to increase with increasing channel width \citep{Davis_2014}, we expect the SMBH mass to be large enough (based on estimates from the $M_{\text{BH}}$ -- $\sigma_\text{e}$ relation) for this not to be an issue here. The resulting data cube has a synthesised beam of $0.20$~arcsec~$\times$~$0.18$~arcsec ($\approx26\times23$~pc$^2$), sampled with $0.04$~arcsec spaxels (spatial pixels), and a RMS noise of $0.44$~mJy~beam$^{-1}$~channel$^{-1}$. Here and throughout the paper the beam sizes are all full-widths at half-maxima (FWHM). The properties of our adopted data cube are summarised in Table~\ref{tab:CO}.

\begin{table}
    \centering
    \renewcommand\thetable{1}
    \caption{CO data cube properties.}
    \begin{tabular}{lc}
        \hline
         Property & Value \\
         \hline
         Spatial extent (pix) & $800\times800$ \\
         Spatial extent (arcsec) & $32.0\times32.0$ \\
         Spatial extent (kpc) & $4.2\times4.2$\\
         Pixel scale (arcsec~pix$^{-1}$) & $0.04$\\
         Pixel scale (pc~pix$^{-1}$) & $5.2$\\
         Velocity range (km~s$^{-1}$) & $1300$ -- $2800$\\
         Channel width (km~s$^{-1}$) & $30$\\
         RMS noise (mJy~beam$^{-1}$~channel$^{-1}$) & $0.44$\\
         Number of constraints & $86,091$ \\
         Synthesised beam (arcsec) & $0.20\times0.18$\\
         Synthesised beam (pc) & $26\times23$\\
         \hline 
    \end{tabular}
    \label{tab:CO}
\end{table}

\begin{figure*}
  \centering
  \includegraphics[scale=0.46]{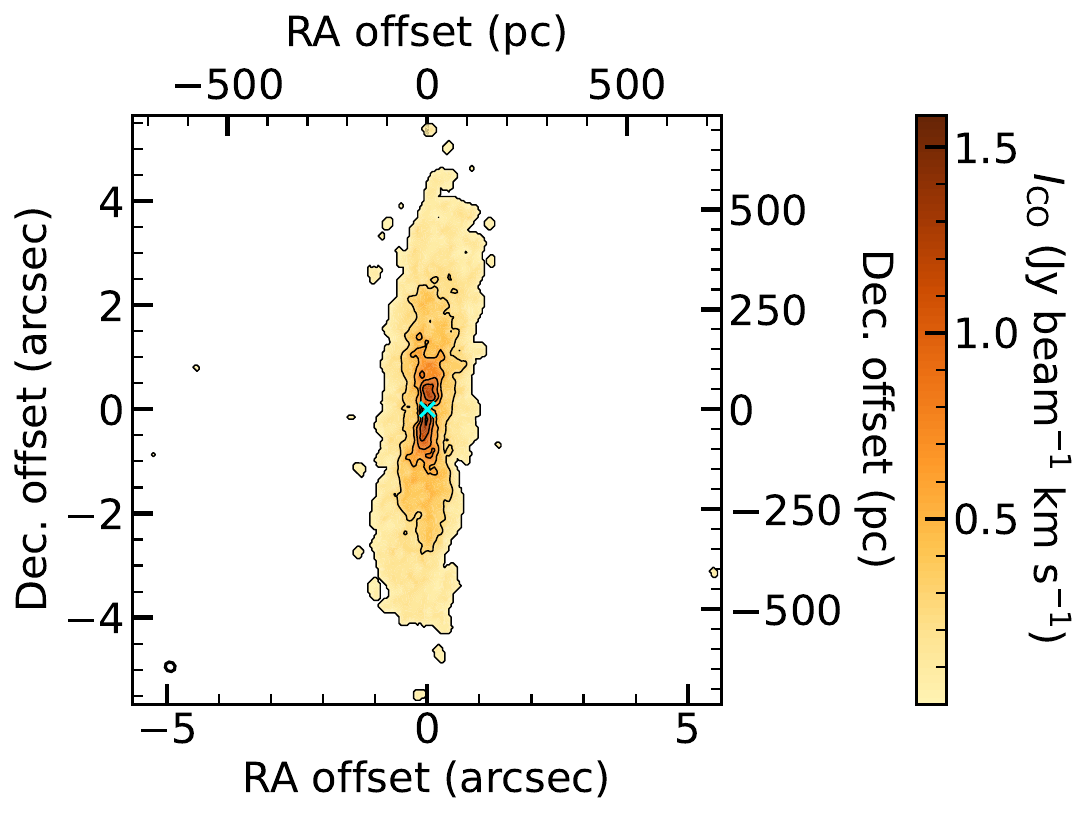}
  \hspace{0.4cm}
  \includegraphics[scale=0.46]{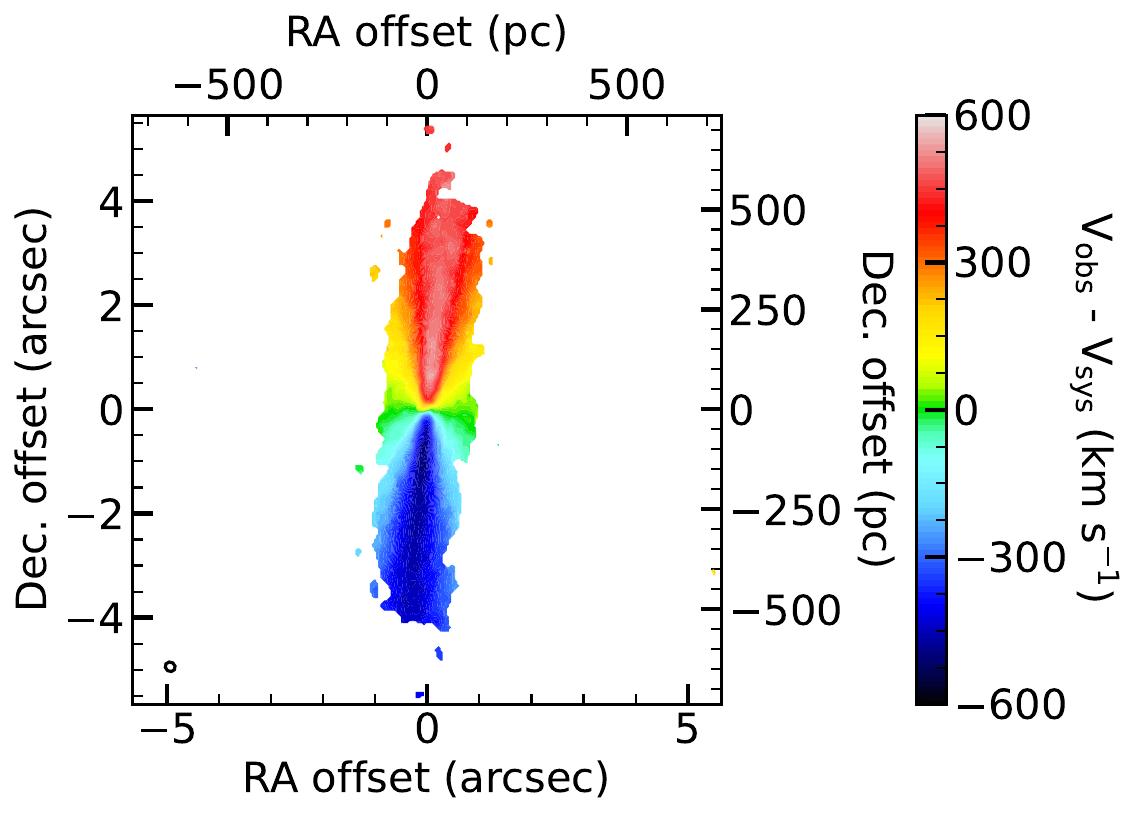}
  \newline
  \hspace*{-1.5cm}
  \includegraphics[scale=0.46]{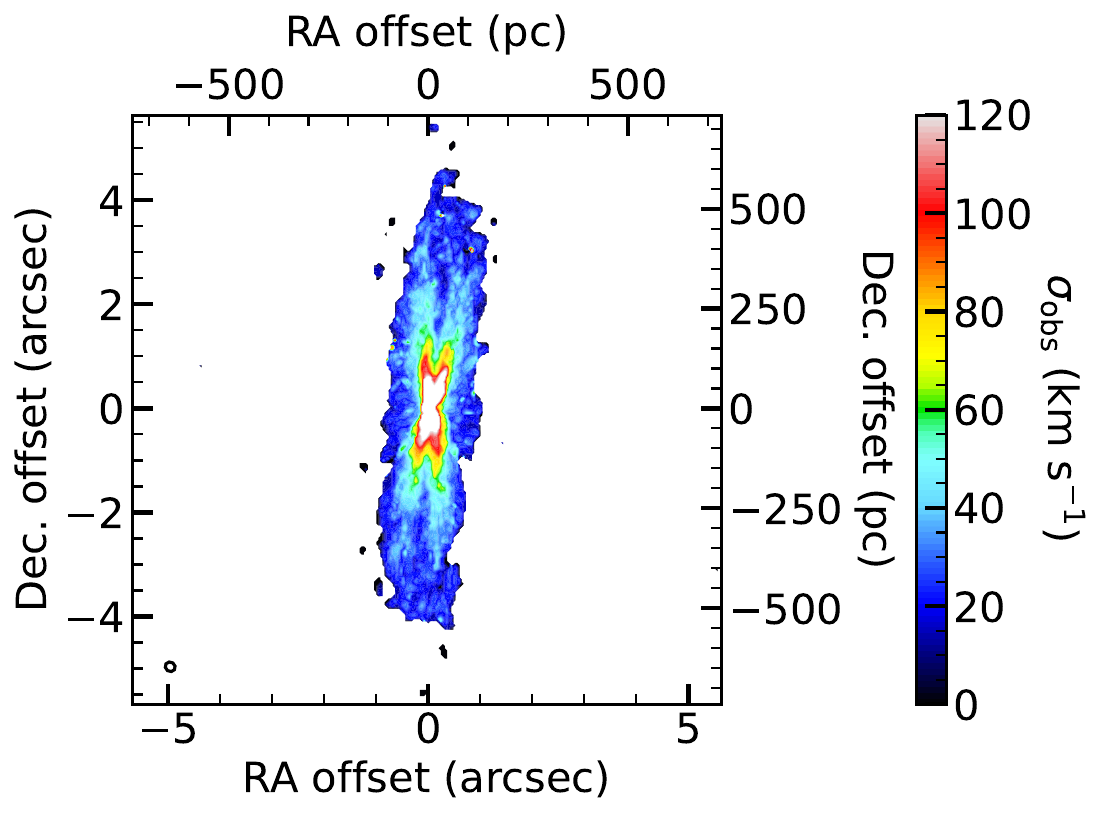}
  \hspace{0.15cm}
  \includegraphics[scale=0.41]{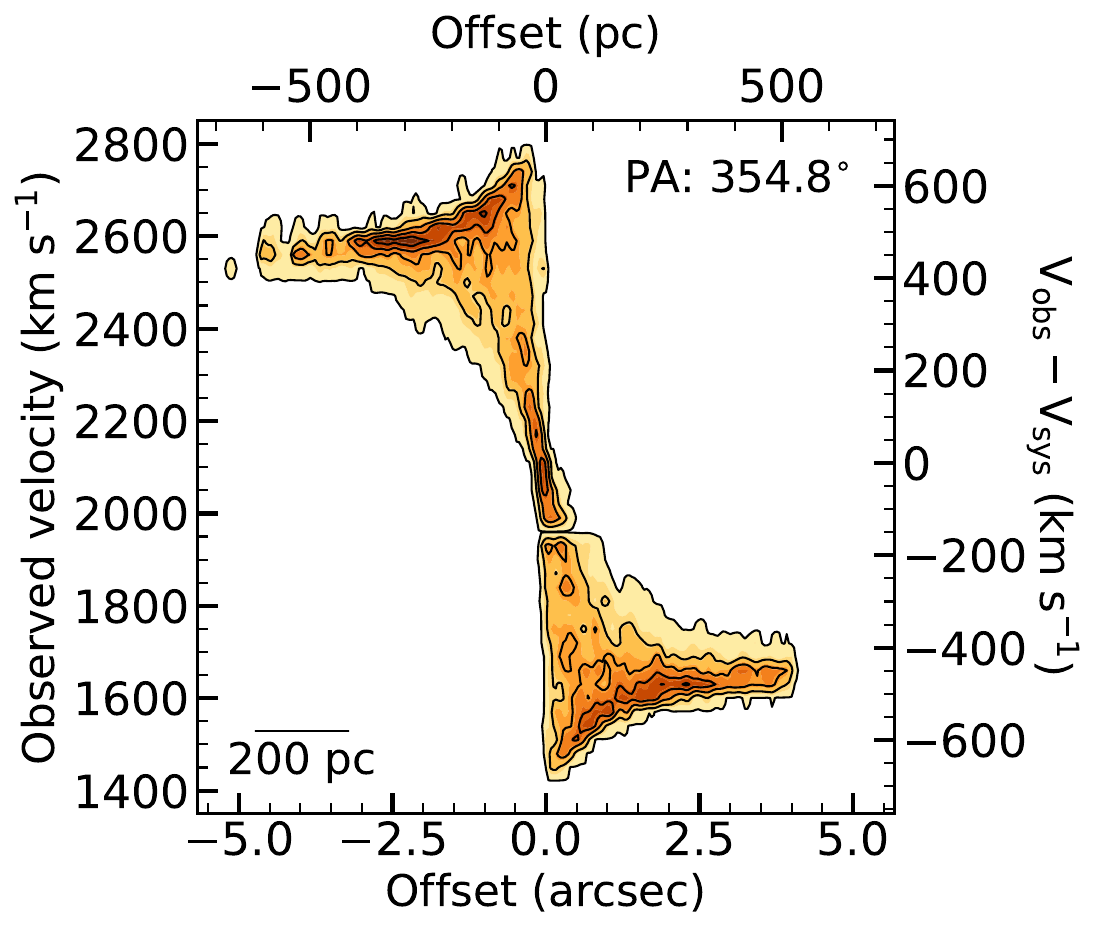}
  \caption{$^{12}$CO(3–2) data products of NGC~4751 derived from our ALMA data. \textbf{Top-left:} zeroth-moment (integrated-intensity) map, with a cyan cross overlaid indicating the dynamical centre of the galaxy. \textbf{Top-right:} first-moment (intensity-weighted mean line-of-sight velocity) map. \textbf{Bottom-left:} second-moment (intensity-weighted line-of-sight velocity dispersion) map. \textbf{Bottom-right:} kinematic major-axis position-velocity diagram. The synthesised beam is shown in the bottom-left corner of each map as an open ellipse. A scale bar is shown in the bottom-left corner of the PVD.}
  \label{fig:mom012}
\end{figure*}

Integrated-intensity, intensity-weighted mean line-of-sight velocity and intensity-weighted line-of-sight velocity dispersion maps were created using a standard masked-moment technique \citep{Dame_2011} implemented in the \textsc{pymakeplots} package\footnote{\url{https://github.com/TimothyADavis/pymakeplots}}. The CO data cube without primary-beam correction was first spatially and spectrally smoothed, with a boxcar spatial kernel of $1.5$ times the width of the synthesised beam and a boxcar spectral kernel of $4$ times the channel width. This smoothed data cube was then clipped at a threshold of $5$ times the RMS noise of its line-free channels, creating a binary mask. We then applied this mask to the unsmoothed primary beam-corrected (i.e.\ the original) data cube to create the moment maps. A kinematic major-axis position-velocity diagram (PVD) was created by taking a cut through the primary beam-corrected cube (with the same mask applied) at a position angle of $354\fdg8$. The resulting moment maps and PVD are shown in Fig.~\ref{fig:mom012}. 

We detect a highly-inclined disc of molecular gas, $\approx9$~arcsec~$\times$~$2$~arcsec in projection, in regular rotation and coincident with but much smaller than the dust disc (see Fig.~\ref{fig:HST}). The molecular gas of NGC~4751 appears to extends all the way to the centre, with no visible circumnuclear hole nor clear depression at the dynamically-determined centre of the galaxy (marked with a cyan cross in Fig.~\ref{fig:mom012}). There is however a small depression $\approx0.2$~arcsec north of the centre. The vast majority of existing molecular gas SMBH measurements used lower-$J$ transitions, most commonly $^{12}$CO(2–1). The lack of a central hole in our $^{12}$CO(3–2) data could suggest that thermal effects in accretion discs reduce the fraction of gas emitting in the low-$J$ transitions in the innermost regions. The PVD shows a rotation curve with a remarkably extended Keplerian rise in the centre.

\begin{figure}
    \centering    
    \includegraphics[scale=0.079,trim={5cm 0 0 0}]{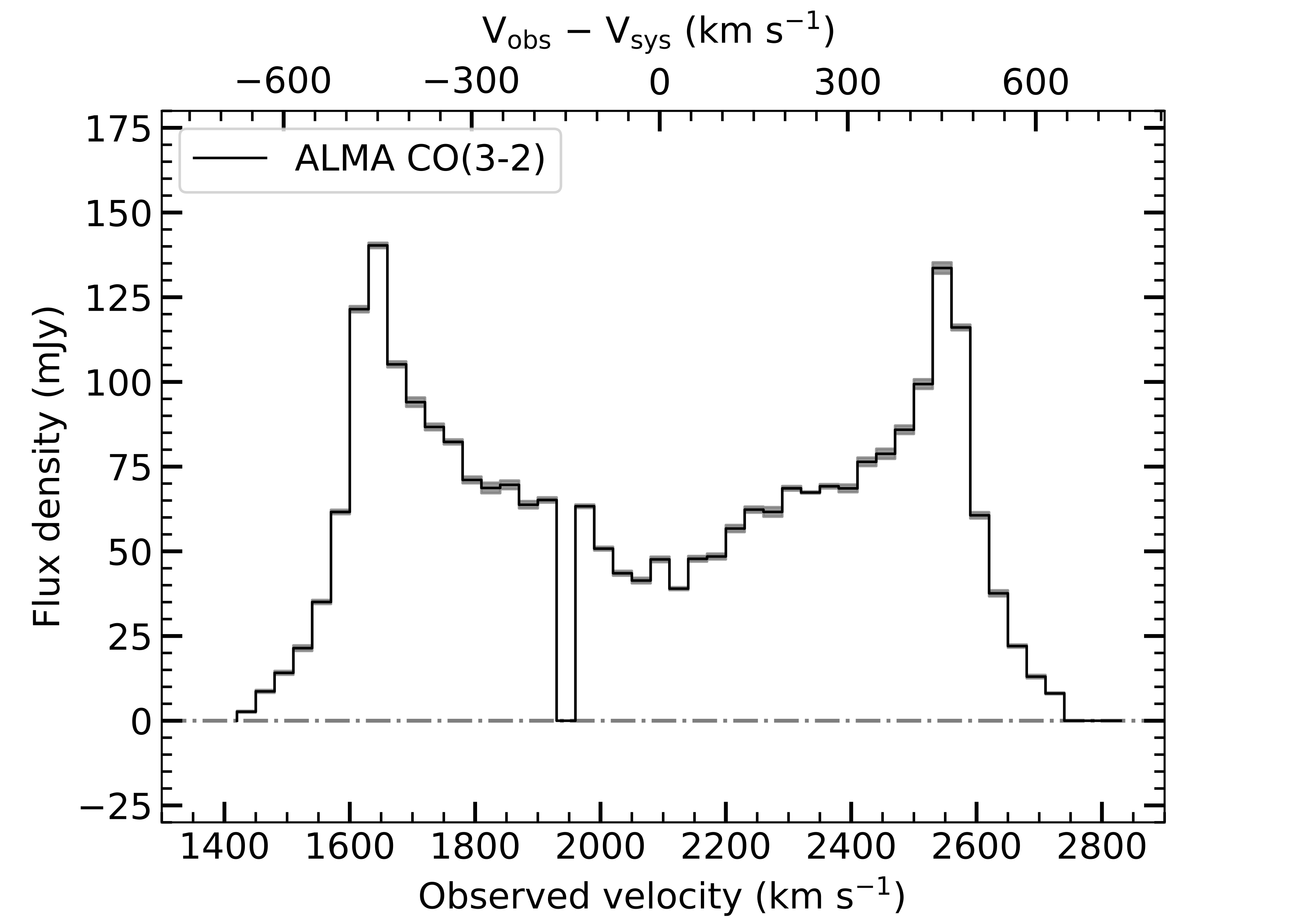}
    \caption{$^{12}$CO(3–2) integrated spectrum of NGC~4751, with the characteristic double-horn shape of a rotating disc. Statistical uncertainties are shown as grey shading. The spectrum was extracted from a $12$~arcsec~$\times$~$12$~arcsec ($\approx1.6\times1.6$~kpc$^2$) region centred on the galaxy centre, that includes all detected emission. The data were obtained in two spectral windows which do overlap, but the pipeline calibration process masked the end channels, likely due to bandpass calibration issues, hence the small gap at $\approx1940$~km~s$^{-1}$.}
    \label{fig:spectrum}
\end{figure}

Figure~\ref{fig:spectrum} shows the $^{12}$CO(3–2) integrated spectrum of NGC~4751, extracted from a $12$~arcsec~$\times$~$12$~arcsec ($\approx1.6\times1.6$~kpc$^2$) region around the centre of the galaxy, using the primary beam-corrected cube with the same mask applied as for the moment maps and PVD. The spectrum  exhibits the characteristic double-horn shape of a rotating disc. The integrated flux of our cube (measured within the mask defined above) is $80.5\pm0.2\,\text{[stat]}\pm8.1\,\text{[sys]}$~Jy~km~s$^{-1}$, where the systematic uncertainty represents the $10$~per~cent absolute flux calibration accuracy \citep{ALMA_Cycle4}. The statistical uncertainty is calculated as in \citet{Liang_2023}, where for each line channel we calculate the integrated flux uncertainty within the two-dimensional (2D) masked region and then propagate this uncertainty in the standard manner to the flux integrated over all channels. We assume no covariance between adjacent channels due to the large channel width. This number includes emission in the missing channels, which was interpolated linearly from the closest channels on either side. Assuming a $^{12}$CO(3–2)/$^{12}$CO(1–0) line ratio of $0.31$ (in brightness temperature units; \citealt{Leroy_2022}) and a standard CO(1-0)-to-molecule conversion factor (with contribution from heavy elements included) $\alpha_\text{CO(1–0)}=4.35$~M$_\odot$~pc$^{-2}$~(K~km~s$^{-1}$)$^{-1}$, we obtain a CO(3-2)-to-molecule conversion factor $\alpha_\text{CO(3–2)}=14.0$~M$_\odot$~pc$^{-2}$~(K~km~s$^{-1}$)$^{-1}$, yielding a total molecular gas mass of $(2.27\pm0.01[\text{stat}]\pm0.22[\text{sys}])\times10^8$~M$_\odot$ over an inner region of $\approx1$~kpc in diameter.

\subsection{Continuum emission}

We imaged the continuum emission of NGC~4751 using the same continuum spectral windows and line-free channels of the line spectral windows that were used for continuum subtraction in Section~\ref{section:line emission}. We used the multi-frequency synthesis mode of the \textsc{casa} task \texttt{tclean} and Briggs weighting with a robust parameter of $0.5$, resulting in a continuum emission map with a central frequency of $343.548$~GHz ($0.87$~mm) and a RMS noise of $0.09$~mJy~beam$^{-1}$. We cleaned this map to a threshold equal to $3$ times the RMS noise and detected a small central source. Fitting this source with a 2D Gaussian function using the \textsc{casa} task \texttt{imfit} reveals a spatially-resolved source with deconvolved full-widths at half-maximum (FWHM) of $(106\pm8)\times(75\pm9)$~marcsec$^2$ and an integrated flux density of $9.9\pm0.2$~mJy. The properties of the continuum emission are listed in Table~\ref{tab:continuum}.

\begin{table}
    \centering
    \renewcommand\thetable{2}
    \caption{Parameters of the NGC~4751 continuum image and the detected cleaned continuum source.}
    \begin{tabular}{lc}
        \hline
         Image property & Value \\
         \hline
         Image (pix) & $800\times800$ \\
         Image (arcsec) & $32\times32$ \\
         Image size (kpc) & $4.2\times4.2$\\
         Pixel scale (arcsec~pix$^{-1}$) & $0.04$ \\
         Pixel scale (pc~pix$^{-1}$) & $5.2$ \\
         RMS noise (mJy beam$^{-1}$) & $0.09$ \\
         Synthesised beam (arcsec) & $0.19\times0.17$ \\
         Synthesised beam (pc) & $25\times22$ \\
         \hline
         Source property & Value \\ 
         \hline 
         Right ascension (J2000.0) & $12^{\text{h}}52^{\text{m}}50\dotes7444\pm0\dotes0001$ \\
         Declination (J2000.0) & $-42\degree39\arcmin35\farcs546\pm0\farcs001$ \\
         Integrated flux (mJy) & $9.9\pm0.2$ \\
         Deconvolved size (mas) & $(106\pm8)\times(75\pm9)$ \\
         Deconvolved size (pc) & $(14\pm1)\times(10\pm1)$ \\
         \hline  
    \end{tabular}
    \label{tab:continuum}
\end{table}


\section{Dynamical Modelling of the CO gas kinematics}
\label{section:Dynamical Modelling}

\subsection{Modelling method}
\label{section: Modelling process}

Dynamical modelling is carried out by fitting a model to the observed molecular gas distribution and kinematics, i.e.\ the data cube discussed in the previous section. This method of SMBH mass determination has been used in other papers of this series \citep[e.g.][]{Davis2017, Smith_2021, Ruffa_2023}.

First we need to model the molecular gas distribution of the galaxy. Due to its simple morphology, we can reproduce the molecular gas distribution of NGC~4751 with an infinitely-thin (i.e.\ 2D) axisymmetric exponential surface brightness profile parameterised as 
\begin{equation}
    I(R)\propto e^{-R/R_0}\,\,\,,
    \label{eq:surface density}
\end{equation}
where $R$ is the galactocentric radius and $R_0$ the exponential scale length, the latter being a free parameter of our model. This exponential surface brightness profile matches the NGC~4751 data well, and the choice of a centrally-concentrated profile has also been implemented in other works \citep[e.g.][]{Davis2013a, Ruffa_2019}. However it is worth noting that such a profile may not be generally appropriate for all ETGs. We also conducted tests where we allowed the exponential surface brightness profile to have a central hole, but the results of these tests suggest that there is no hole or that the hole is significantly smaller than our synthesised beam size. 

Whilst no clear hole is visible in the moment maps, given the high inclination of the molecular gas disc it is of course very hard to rule out the presence of a hole of a size similar to or smaller than the synthesised beam. Indeed, at such a high inclination a small central hole would not be well resolved along the minor axis and emission along the minor axis of the galaxy might be beam-smeared and give the impression of a filled central hole. Nevertheless, due to the modelling results above, we proceed assuming no hole is present. Our surface density profile is then scaled to match the integrated $^{12}$CO(3–2) flux, leading to another free parameter of our model.

Another free parameter of our model is the velocity dispersion of the molecular gas ($\sigma_{\text{gas}}$), which we assume to be constant at all radii and relatively small compared to the rotational velocities of the molecular gas disc. This is a standard assumption when modelling nearby, dynamically-cold molecular gas discs (e.g.\ \citealt{Davis2017}, \citealt{WISDOM_V}, \citealt{Dominiak_2024}). 

Second, to estimate the SMBH mass accurately, we must account for the stellar mass contribution to the molecular gas kinematics. In principle the contribution dark matter should also be considered, but at the small radii probed here that contribution is expected to be negligible (it would in any case be largely degenerate with that of the stars; \citealt{Cappellari_2013a}). We parameterise the stellar light distribution using a multi-Gaussian expansion (MGE) model of a \textit{HST} Wide Field Camera~3 (WFC3) F160W filter ($H$-band) pipeline-drizzled image (GO-15909, PI: Boizelle) with a total exposure time of $997$~s, obtained from the Mikulski Archive for Space Telescopes. For this, we use the \texttt{mge\_fit\_sectors} procedure in the \textsc{MgeFit} package\footnote{\url{https://pypi.org/project/mgefit/}} \citep{Cappellari2002}. This is the longest wavelength \textit{HST} image available, to minimise dust extinction. We adopt the spatial point-spread function (PSF) of the F160W filter from \citet{Dominiak_2024}.

\begin{figure}
    \centering
    \includegraphics[scale = 0.55]{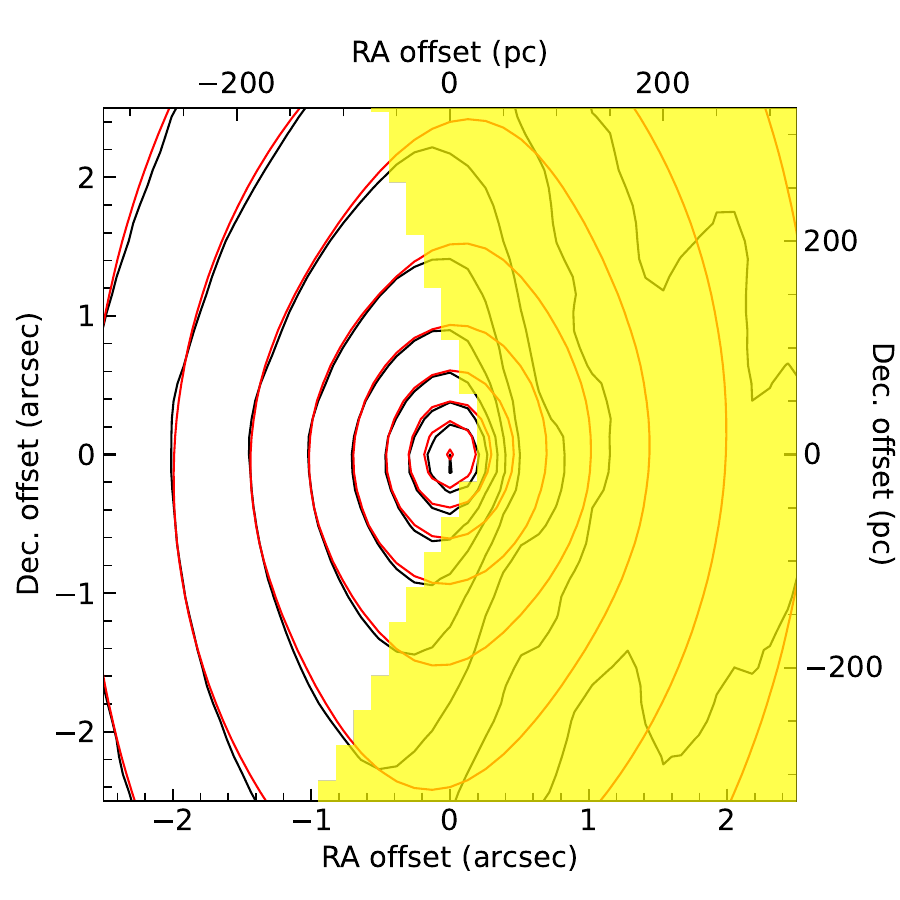}
    \caption{\textit{HST} WFC3 F160W filter image of NGC~4751 (black contours), overlaid with our best-fitting MGE model (red contours). The western side of the image is masked (yellow shading), to exclude the most obviously dust-obscured pixels along the near side of the embedded disc.}
    \label{fig:MGE contours}
\end{figure}

The 2D projection of the stellar light distribution captured by the \textit{HST} image is parameterised by the MGE as a sum of Gaussians, each with a central surface brightness $I^\prime$, a width $\sigma^\prime$ and an axial ratio $q^\prime$, which once convolved by the PSF best reproduce the image. The surface brightnesses can be converted to luminosities using the AB magnitude system with a zero-point of $25.94$~mag \citep{WFC3} and a Solar absolute magnitude in the F160W filter of $4.60$~mag \citep{Willmer2018}. Additionally, we adopt a Galactic extinction of $0.062$~mag from the NASA/IPAC Extragalactic Database \citep{Schlafly_2011} to correct for interstellar reddening. The parameters of these "PSF-deconvolved", projected best-fitting MGE Gaussians are listed in Table~\ref{tab:MGE components} in physical units and the best fit is shown in Fig.~\ref{fig:MGE contours}, revealing significant dust extinction on the near side of the NGC~4751 disc. For our best-fitting model, we masked these dust lanes by hand, thus excluding the most obviously dust-obscured pixels, to minimise their impact on our final model. However, we explore the use of a more complex pixel-by-pixel colour correction of the dust in Section~\ref{subsection:uncertainties}. The resulting MGE model provides a good fit, especially in the central regions the galaxy, which is crucial for determining an accurate SMBH mass. 

\begin{table}
    \renewcommand\thetable{3}
	\centering
	\caption{Parameters of the "deconvolved" best-fitting MGE components.}
	\label{tab:MGE components}
	\begin{tabular}{ccc} 
	\hline
	$\log I_{\odot}^\prime$ & $\log \sigma^\prime$ & $q^\prime$ \\
        (L$_{\odot, \text{F160W}}$~pc$^{−2}$) & (arcsec) & \\
        (1) & (2) & (3) \\
	\hline
	5.491 & $-0.815$ & 0.647 \\
	4.851 & $-0.422$ & 0.494 \\
	4.441 & $ -0.238$ & 0.990 \\
    4.150 & $\phantom{-}0.171$ & 0.750 \\
    4.099 & $\phantom{-}0.177$ & 0.319 \\
    4.002 & $\phantom{-}0.524$ & 0.517 \\
    3.511 & $\phantom{-}0.934$ & 0.397 \\
    2.913 & $\phantom{-}1.411$ & 0.420 \\      
	\hline
    \end{tabular}
    \begin{tablenotes}
          \item \textit{Notes.} "Deconvolved" MGE Gaussian components. (1) Central surface brightness. (2) Standard deviation. (3) Axial ratio.
    \end{tablenotes}
\end{table}

By assuming an inclination and axisymmetry, the 2D Gaussians can be analytically deprojected into three-dimensional (3D) Gaussians. By multiplying each Gaussian by a stellar $M/L$, we can in turn convert the resulting 3D light volume density distribution into a 3D stellar mass volume density distribution, as required for dynamical modelling. This $M/L$ is another free parameter of our model and is usually assumed to be spatially constant. However, in this paper we also consider a $M/L$ varying linearly with radius, as the $J-H$ colour map of NGC~4751 suggests an increase of the colour towards the centre of the galaxy that cannot be explained by an embedded dust disc (see Section~\ref{subsection:uncertainties} for more details).

We implement this radially linearly varying $M/L$ by first deriving the circular velocity curve of our stellar mass model assuming a $M/L$ of unity ($1$~M$_\odot/\text{L}_{\odot,H}$) and using the \texttt{mge\_vcirc} routine of the \textsc{Jeans Anisotropic Modelling} Python package (\textsc{JamPy}\footnote{\url{https://pypi.org/project/jampy}}; \citealt{Cappellari2008,Cappellari_2020}). We then scale these circular velocities by the square root of the desired $M/L$ profile (as circular velocities are proportional to the square root of the enclosed masses). In this linearly-varying $M/L$ case, the $M/L$ intercept {\it and} $M/L$ gradient are thus free parameters of our model. Lastly, we add the contribution of a (free) SMBH mass to the circular velocity curve.

By combining the molecular gas distribution and the circular velocity curve obtained above, the \textsc{Kinematic Molecular Simulation} (\textsc{KinMS}; \citealt{Davis2013a}) task \texttt{KinMS\_fitter}\footnote{\url{https://github.com/TimothyADavis/KinMS_fitter}} recreates the molecular gas disc by simulating it as a set of point particles.
A model data cube is then created by computing the line-of-sight projections of the particles, whilst taking into consideration other "nuisance" free parameters of our model: central position, systemic velocity, inclination and position angle (the latter two assumed to be radially constant). To replicate instrumental effects, this data cube is spatially convolved with the clean Gaussian synthesised beam and then binned (spatially and spectrally) into pixels identical to those of the real data cube.

We use the Gibbs sampler with adaptive stepping \textsc{GAStimator}\footnote{\url{https://github.com/TimothyADavis/GAStimator}} to compare our model and ALMA data cubes. Initially, the Monte Carlo Markov chain (MCMC) algorithm samples the parameter space of all the free parameters of the model, and the step size between each fit is adaptively scaled until the chain converges. Approximately $10$~per~cent of the total number of steps is used to identify the convergent chain. Once this chain is identified and the MCMC has converged, the maximum step size is fixed and the MCMC continues sampling the parameter space, producing samples from the final posterior probability distribution. 

The parameter space is bounded by a set of priors. Some are set manually to ensure a finite converging time; others are allowed to span their entire possible ranges. The priors for all the parameters are linear, except for that on the SMBH mass which is logarithmic due to its large possible dynamic range. Assuming these maximally-ignorant priors and constant Gaussian uncertainties throughout the cube, the posterior probability distribution of a model is proportional to the log-likelihood function $\ln P\propto-0.5\chi^2$, where $\chi^2$ is the sum of the differences between model and data squared normalised by the uncertainties squared. The sum is taken over the mask defined by \textsc{KinMS}, which contains all the pixels with fluxes at least $1.5$ times the RMS noise. To ensure that the large number of pixels within the mask $N$ (listed in Table~\ref{tab:CO}) does not lead to unrealistically small formal uncertainties, we rescale the uncertainties in a standard manner by a factor of $(2N)^{0.25}$. This approach was originally proposed by \citet{vandenBosch2009} for use in least-squares fitting and adapted to the Bayesian framework by \citet{Mitzkus2017}. It is not statistically rigorous, but tries to approximately account for the fact that systematic uncertainties tend to dominate the total uncertainties when fitting a large number of measurements (here tens of thousands). It relies on the sensible but crude assumption that each random uncertainty has an associated systematic one of similar magnitude. In the case of SMBH mass determinations from CO, the method has been shown to generally yield uncertainties that are consistent with more realistic estimates of the systematic uncertainties \citep[]{WISDOM_IV}. Nevertheless, a more thorough exploration of systematic uncertainties is presented in Section~\ref{subsection:uncertainties}.

\begin{figure*}
    \centering
    \includegraphics[scale=0.48, trim={0.8cm 0 0 0}]{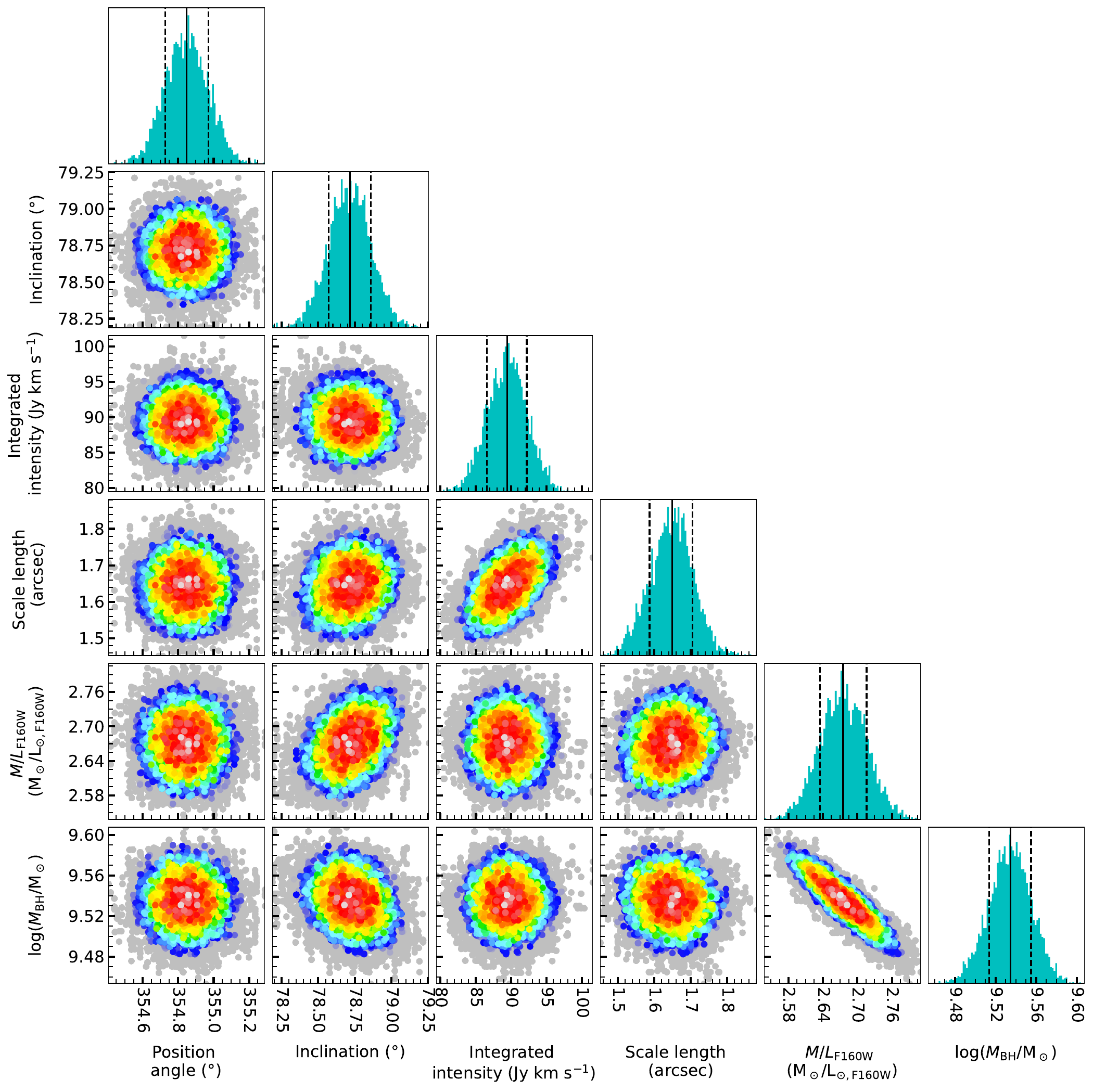}
    \caption{Corner plots of NGC~4751, showing the covariances between selected (primarily non-nuisance) parameters of the constant $M/L$ model. Each data point is one realisation of our model, colour-coded to show the relative log-likelihood of that realisation, with white data points being most likely and blue data points least likely. Coloured data points show models with $\Delta\chi^2<\sqrt{2N}$ of the best-fitting model; grey data points show all remaining models. The histograms show the 1D marginalised posterior distribution of each model parameter; in each case the black solid line indicates the median and the two black dashed lines encompass the $68\%$ confidence interval.}
    \label{fig:CornerPlot}
\end{figure*}

\begin{figure*}
    \centering
    \includegraphics[scale=0.44, trim={0.8cm 0 0 0}]{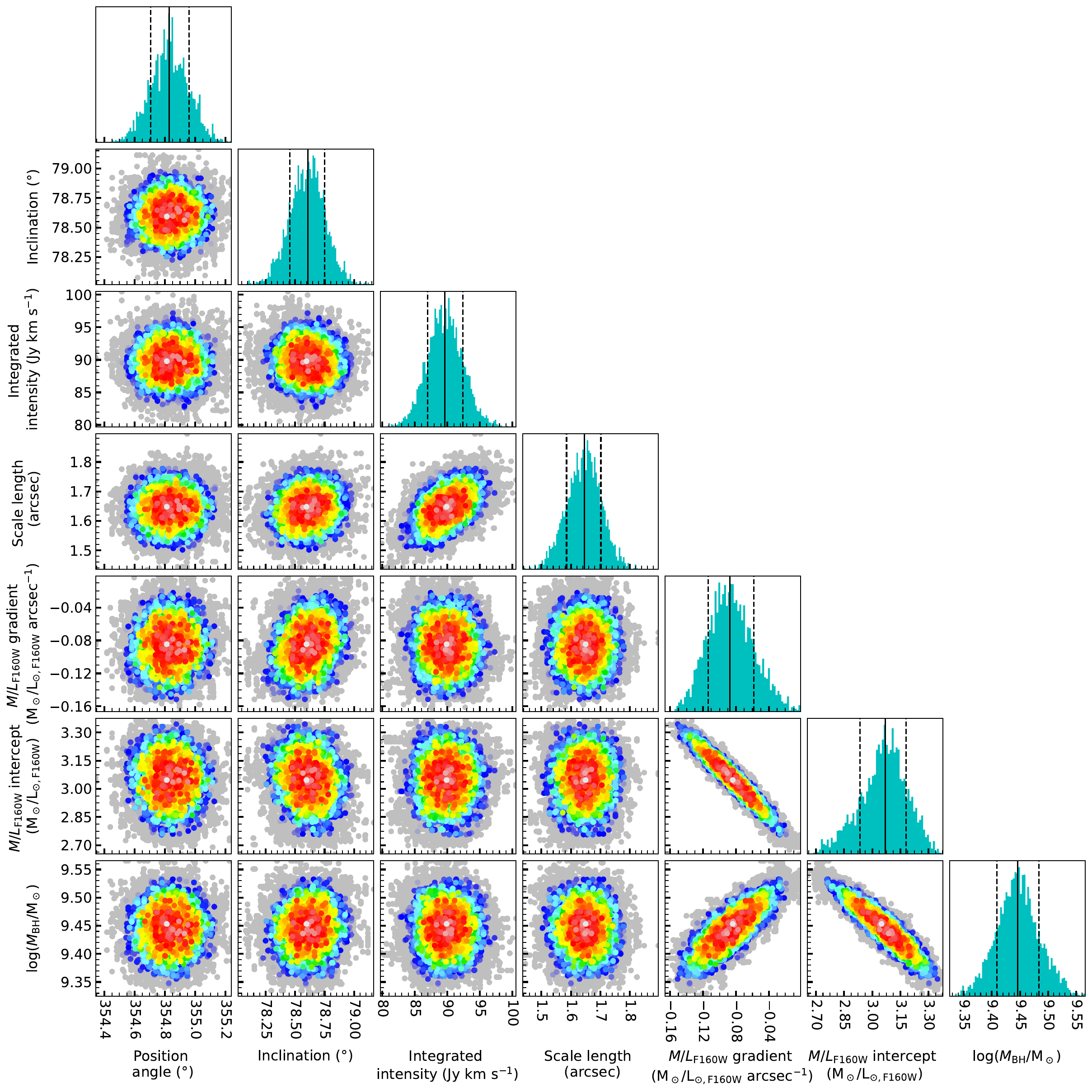}
    \caption{As Fig.~\ref{fig:CornerPlot}, but for the model with a radially linearly-varying $M/L$.}
    \label{fig:CornerPlot_linear}
\end{figure*}

In Figs.~\ref{fig:CornerPlot} and \ref{fig:CornerPlot_linear}, the marginal distributions of our model parameters are shown, produced as a part of the MCMC process. Each data point in the 2D marginalisations represents the log-likelihood of a given model, where the white and red data points are most likely and the blue data points least likely. The one-dimensional (1D) marginalisations are shown in the form of histograms with roughly Gaussian shapes, representing the probability distributions of individual model parameters. 

\subsection{Results}
\label{section:Results}

Throughout our modelling, we considered models with various $M/L$ profiles. The spatially-constant and radially linearly-varying $M/L$ models yield equally good fits, so the results of both are presented below. 

\subsubsection{Model with a constant $M/L$}

This model has a total of $10$ free parameters: SMBH mass, spatially-constant F160W filter stellar mass-to-light ratio ($M/L_\text{F160W}$), molecular gas exponential disc scale length ($R_0$), integrated flux and velocity dispersion ($\sigma_{\text{gas}}$), and the "nuisance" disc parameters position angle, inclination ($i$), systemic velocity ($V_\text{sys}$) and centre position (offsets in right ascension and in declination from the image phase centre). The model parameters and their search ranges, best-fitting values and $1\sigma$ ($68.3$~per~cent) and $3\sigma$ ($99.7$~per~cent) confidence level uncertainties are listed in Table~\ref{tab:Best Fit}.

\begin{table*}
    \begin{threeparttable}
    \renewcommand\thetable{4}
	\centering
	\caption{Best-fitting molecular gas model parameters and associated uncertainties.}
	\label{tab:Best Fit}
	\begin{tabular}{lrcrcrr}
		\hline
	Parameter & & Prior & & Best fit & $1\sigma$           uncertainty
        & $3\sigma$ uncertainty \\
        \hline
        \multicolumn{7}{c}{\textbf{Model with a constant $M/L$}}\\
        \hline
        \textbf{Mass model} \\
        $\log(M_\text{BH}/\text{M}_\odot)$ & $7$ & $\rightarrow$ & $11$ & $9.54$ & $\pm0.02$ & $-0.06,+0.05$ \\
        Stellar $M/L_{\text{F160W}}$ (M$_\odot$/L$_{\odot,\text{F160W}}$) & $0.0$ & $\rightarrow$ & $5.0$ & $2.68$ & $\pm0.04$ & $\pm0.11$ \\
        \\
        \textbf{Molecular gas disc} \\
        Velocity dispersion (km~s$^{-1}$) & $0.0$ & $\rightarrow$ & $50.0$ & $15.1$ & $\pm2.3$ & $-6.7,+6.5$ \\
        Scale length (arcsec) & $0.0$ & $\rightarrow$ & $5.0$ & $1.65$ & $\pm0.06$ & $-0.15,+0.16$ \\
        Integrated intensity (Jy~km~s$^{-1}$) & $0.0$ & $\rightarrow$ & $150.0$ & $89.5$ & $-2.9, +2.8$ & $-7.8,+7.5$ \\
        \\ 
        \textbf{Nuisance parameters} \\
        Centre RA offset (arcsec) & $-5.4$&$\rightarrow$ &$5.4$ & $0.026$ & $\pm 0.004$ &$\pm 0.01$\\
        Centre Dec.\ offset (arcsec) & $-4.0$ &$\rightarrow$ &$4.0$ & $0.31$ & $\pm 0.01$ & $\pm0.02$\\
        Inclination ($\degr$) & $71.4$ & $\rightarrow$ & $89.9$ & $78.7$ & $\pm0.1$ & $-0.5, +0.6$ \\
        Position angle ($\degr$) & $0.0 $ & $\rightarrow$ & $359.9$ & $354.8$ & $\pm0.1$ & $-0.3, +0.4$ \\
        Systemic velocity (km~s$^{-1}$) & $1800$ & $\rightarrow$ & $2400$ & $2094.5$ & $\pm1.1$ & $-3.3,+2.9$ \\
	\hline
        \multicolumn{7}{c}{\textbf{Model with a radially linearly-varying $M/L$}}\\
        \hline
        \textbf{Mass model} \\
        $\log(M_\text{BH}/\text{M}_\odot)$ & $7$ & $\rightarrow$ & $11$ & $9.45$ & $\pm0.04$ & $\pm0.10$ \\
        Stellar $M/L_{\text{F160W}}$ intercept (M$_\odot$/L$_{\odot,\text{F160W}}$) & $0.0$ & $\rightarrow$ & $5.0$ & $3.07$ & $-0.13, +0.11$ & $-0.35, +0.27$ \\
        Stellar $M/L_{\text{F160W}}$ gradient (M$_\odot$/L$_{\odot,\text{F160W}}$~arcsec$^{-1}$) & $-2.0$ & $\rightarrow$ & $2.0$ & $-0.09\phantom{-}$ & $\pm0.03$ & $-0.06, +0.08$ \\
        \\
        \textbf{Molecular gas disc} \\
        Velocity dispersion (km~s$^{-1}$) & $0.0$ & $\rightarrow$ & $50.0$ & $14.8$ & $\pm1.5$ & $-4.3,+4.4$ \\
        Scale length (arcsec) & $0.0$ & $\rightarrow$ & $5.0$ & $1.65$ & $\pm0.06$ & $-0.16,+0.15$ \\
        Integrated intensity (Jy~km~s$^{-1}$) & $0.0$ & $\rightarrow$ & $150.0$ & $89.6$ & $-2.7, +2.8$ & $-7.3,+7.8$ \\
        \\ 
        \textbf{Nuisance parameters} \\
        Centre RA offset (arcsec) & $-5.4$&$\rightarrow$ &$5.4$ & $0.025$ & $\pm 0.004$ &$\pm 0.01$\\
        Centre Dec.\ offset (arcsec) & $-4.0$ &$\rightarrow$ &$4.0$ & $0.31$ & $\pm 0.01$ & $\pm0.02$\\
        Inclination ($\degr$) & $71.4$ & $\rightarrow$ & $89.9$ & $78.6$ & $-0.2, +0.1$ & $-0.5, +0.4$ \\
        Position angle ($\degr$) & $0.0 $ & $\rightarrow$ & $359.9$ & $354.8$ & $\pm0.1$ & $-0.3, +0.4$ \\
        Systemic velocity (km~s$^{-1}$) & $1800$ & $\rightarrow$ & $2400$ & $2094.3$ & $-1.1, +1.0$ & $-4.4,+4.3$ \\
        \hline
	\end{tabular}
    \begin{tablenotes}[flushleft]
          \item \textit{Note.} The RA and Dec.\ offsets are measured with respect to the image phase centre, $12^{\text{h}}52^{\text{m}}50\dotes7484$, $-42\degree39\arcmin35\dotarc880$ (J2000.0).
    \end{tablenotes}
    \end{threeparttable}
\end{table*}

The final MCMC chain had $200,000$ steps. It is clear that there is a massive dark object at the centre of NGC~4751, with a best-fitting mass of $3.43^{+0.45}_{-0.44}\times10^9$~M$_\odot$, where here and throughout this paper the uncertainties are stated at the $3\sigma$ ($99.7$~per~cent) confidence level. The best-fitting F160W filter $M/L$ is $(2.68\pm0.11)$~M$_\odot$/L$_{\odot,\text{F160W}}$. Figure~\ref{fig:CornerPlot} also shows a degeneracy between $M_{\text{BH}}$ and $M/L_\text{F160W}$, equivalent to the conservation of (total) dynamical mass (see e.g. \citealt{WISDOM_IV} for more details).

\begin{figure*}
    \centering\includegraphics[scale=0.58, trim={3.9cm 0 3cm 0}]{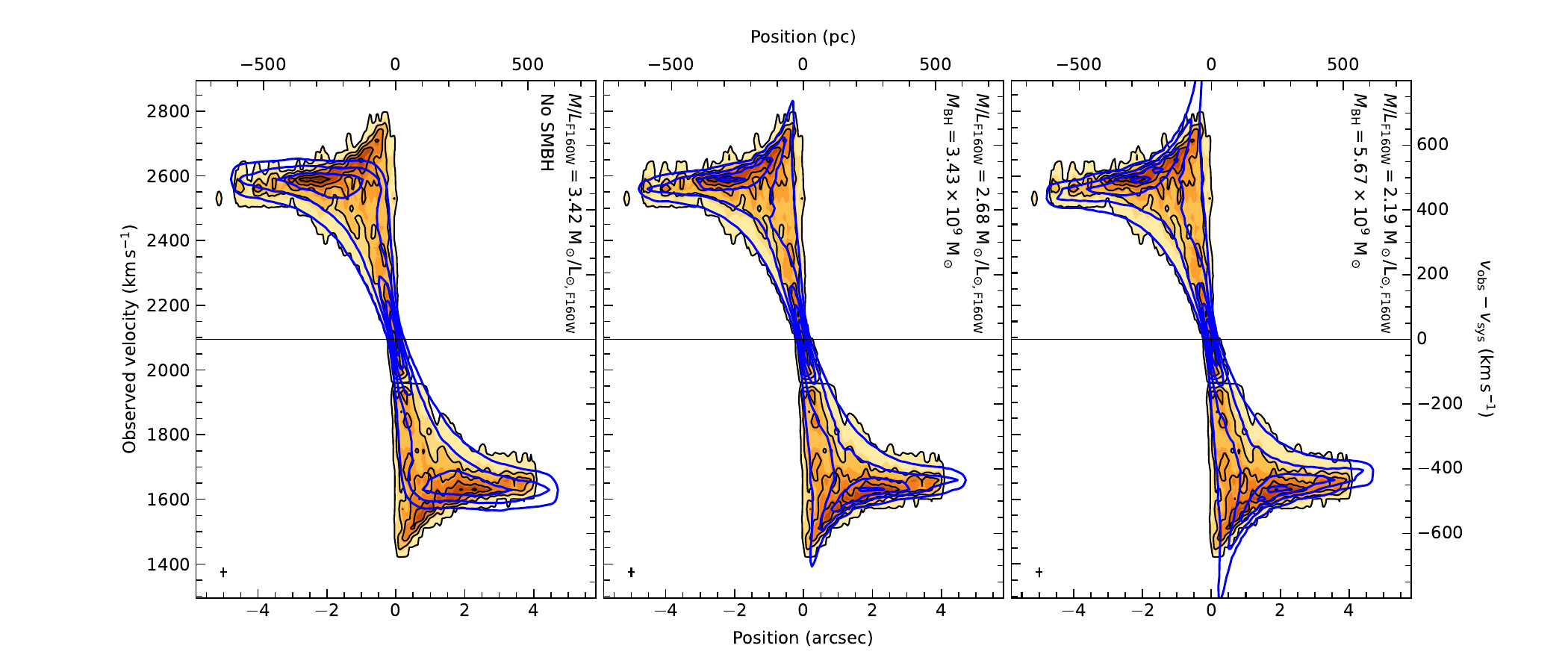}
    \caption{Observed kinematic major-axis PVD of NGC~4751 (orange scale with black contours), overlaid with the best-fitting no SMBH (left), free SMBH mass (centre) and overly-massive (by $\approx0.2$~dex) SMBH (right) model (blue contours), respectively. The SMBH mass and $M/L$ of each model are listed in the top-right corner of each panel. An error bar is shown in the bottom-left corner of each panel, showing the size of the synthesised beam along the kinematic major-axis and the channel width. The need for a central dark mass to fully account for the gas kinematics at all radii is clear.}
    \label{fig:PVD_NGC4751}
\end{figure*}

The quality of our best-fitting model can be gauged from Fig.~\ref{fig:PVD_NGC4751}, where from left to right we overlay the best-fitting model with no SMBH, our best-fitting SMBH and an overly-massive SMBH ($\approx0.2$~dex more massive than the best-fitting SMBH), respectively, over the kinematic major-axis PVD. In the first and third case all parameters other than the SMBH mass were allowed to vary. As expected for the best-fitting no SMBH model, a higher stellar $M/L$ is derived as the model attempts to account for the high rotation velocities at small radii without a SMBH. Nevertheless, the model is unable to reproduce those central velocities without greatly exceeding the relatively low velocities at larger radii. The compromise reached is thus unsatisfactory at both small and large radii. Again as expected, the best-fitting overly-massive SMBH model yields a smaller $M/L$, but the fit is very poor at small radii, the model over-shooting even the highest velocities. The best-fitting SMBH model not only reproduces the data best, but the fit is very good at all radii and velocities.

\subsubsection{Model with a radially linearly-varying $M/L$}

This model has a total of $11$ free parameters: a stellar $M/L$ intercept and a stellar $M/L$ gradient replace the constant $M/L$ of the previous model, while the other $9$ parameters remain the same. The model parameters and their search ranges, best-fitting values and $1\sigma$ and $3\sigma$ confidence level uncertainties are listed in Table~\ref{tab:Best Fit}. 

The final MCMC chain has $200,000$ steps. The model has a best-fitting SMBH mass of $2.79_{-0.57}^{+0.75}\times10^9$~M$_\odot$ and a best-fitting F160W filter $M/L$ of $\left(M/L_\text{F160W}\right)/\left(\text{M}_\odot/\text{L}_{\odot,\text{F160W}}\right)=3.07^{+0.27}_{-0.35}-0.09^{+0.08}_{-0.06}\,\left(R/\text{arcsec}\right)$. ETG $M/L$ gradients have been explored in a number of previous works exploiting molecular gas dynamics \citep[e.g.][]{Davis_McDermid_2016, WISDOM_III, WISDOM_V}. In these works, both positive and negative $M/L$ gradients were detected and in most of these cases the gradients are mild, very much like our results (e.g.\ $-0.37$~M$_\odot$/L$_{\odot,\text{F160W}}$~kpc$^{-1}$ in NGC~524 compared to our $-0.69$~M$_\odot$/L$_{\odot,\text{F160W}}$~kpc$^{-1}$; \citealt{WISDOM_V}). 

Figure~\ref{fig:CornerPlot_linear} again shows a negative correlation between the $M/L$ intercept (i.e.\ the $M/L$ in the centre of the galaxy) and the SMBH mass, again equivalent to the conservation of (total) dynamical mass. Consequently, there is also a positive correlation between SMBH mass and $M/L$ gradient -- as more mass is attributed to the SMBH, the $M/L$ intercept decreases and thus the $M/L$ gradient becomes less negative (to still fit the outer regions of the galaxy well).

\begin{figure*}
    \centering\includegraphics[scale=0.58, trim={3.9cm 0 3cm 0}]{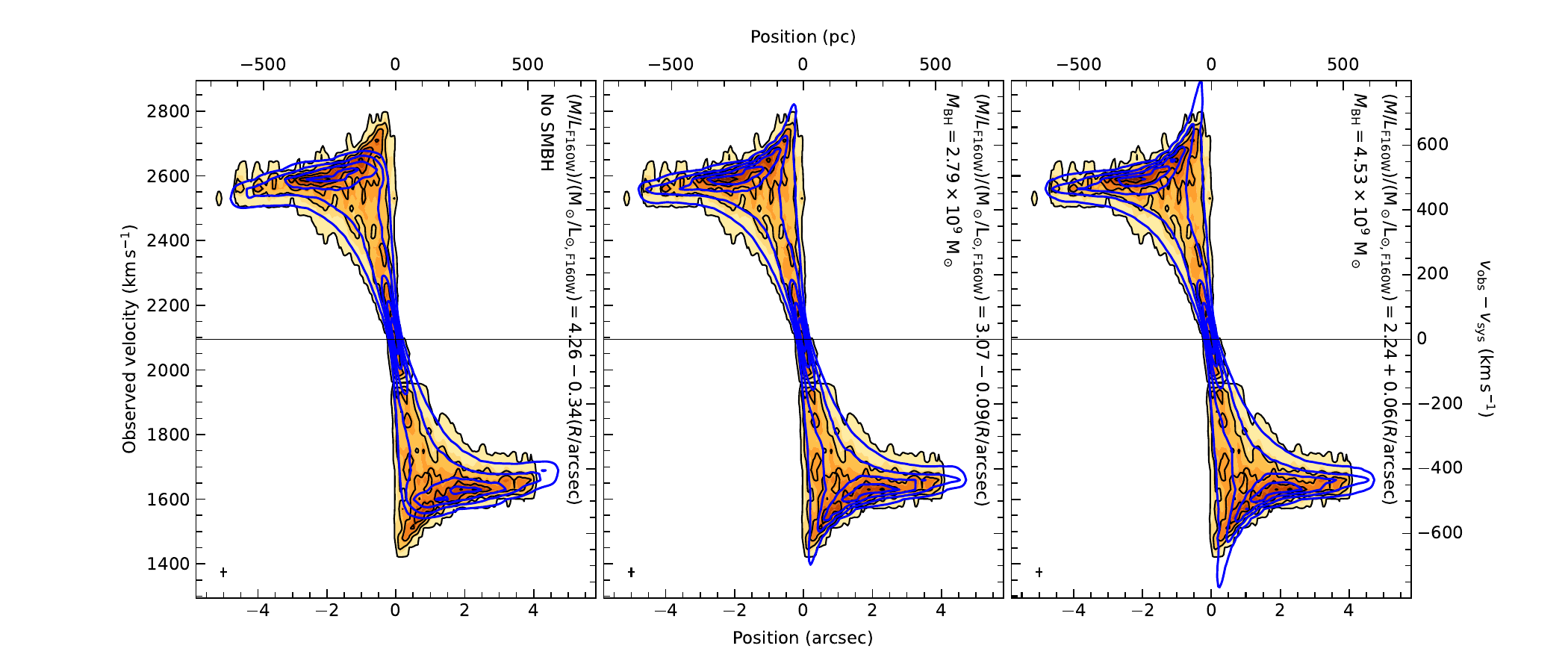}
    \caption{As Fig.~\ref{fig:PVD_NGC4751}, but for models with radially linearly-varying $M/L$.}
    \label{fig:PVD_NGC4751_linearML}
\end{figure*}

As before, the quality of our best-fitting model is easiest to judge by overlaying it over the kinematic major-axis PVD. From left to right in Fig.~\ref{fig:PVD_NGC4751_linearML}, we overlay the best-fitting model with a radially linearly-varying $M/L$ and no-SMBH, a SMBH and an overly-massive SMBH ($\approx0.2$~dex more massive than the best-fitting SMBH), respectively, allowing all parameters other than the SMBH mass to vary in the first and third case. As expected, the best-fitting no SMBH model has a higher $M/L$ intercept, to account for the high rotation velocities at small radii, and a more negative $M/L$ gradient, to lower the overall $M/L$ at larger radii. As before, the model fails to reproduce the extremely high central velocities without also exceeding the low velocities at large radii. The best-fitting overly-massive SMBH model yields a significantly smaller $M/L$ intercept that contributes less dynamical mass to the stellar component. As expected, the $M/L$ gradient is also now positive, as otherwise the low central $M/L$ would lead to undershooting the velocities at large radii. Despite this, the overly-massive SMBH model is unable to reproduce the data satisfactorily. Once again, the best-fitting free SMBH model reproduces the data best, fitting well at all radii and velocities.

\section{JAM modelling of the stellar kinematics}
\label{section:JAM models}

We obtain an alternative SMBH mass estimate using the JAM method \citep{Cappellari2008,Cappellari_2020} and the SINFONI stellar kinematics of NGC~4751 published by \citet{Rusli2013}. The JAM method is particularly well-suited to early-type fast-rotators, of which NGC~4751 is one \citep{Rusli2013}. We first use the \textsc{WebPlotDigitizer} software\footnote{\url{https://github.com/ankitrohatgi/WebPlotDigitizer}} to extract the stellar kinematics (line-of-sight mean velocity $V$ and velocity dispersion $\sigma$) from the published paper (Fig.~24 of \citealt{Rusli2013}). From these, we compute the second velocity moment as $V_\text{rms}\equiv\sqrt{V^2+\sigma^2}$, which can be directly compared to the JAM predictions. 

We fit an image of the SINFONI PSF, kindly provided to us by Jens Thomas, using the \texttt{mge\_fit\_sectors} procedure of the \textsc{MgeFit} package as before. Two circular Gaussians were sufficient to provide a good representation of the PSF, with best-fitting parameters $\mathrm{FWHM}=[0\farcs24, 0\farcs82]$ and $\mathrm{Frac}=[0.55, 0.45]$, where Frac is the fractional light contribution of each PSF component. As expected the FWHM of the narrow component is close to the FWHM of $0\farcs22$ listed in Table~2 of \citet{Rusli2013}, while the FWHM of the broad component is typical of FWHM measured from similar SINFONI observations (see e.g.\ Table~3 of \citealt{Thater2019}).

We adopt the same \textit{HST} WFC3 F160W filter MGE parametrisation used to model the molecular gas kinematics in Section~\ref{section: Modelling process} to describe the stellar surface brightness of the JAM dynamical models. Moreover, we assume that mass follows light and adopt a constant orbital anisotropy. The latter assumption is unlikely to be accurate over large spatial scales, and in fact our more spatially-extended CO model indicates an $M/L$ gradient. However, the JAM models are constrained only by the stellar kinematics at $R\lesssim1\arcsec$, so the results are only weakly sensitive to possible $M/L$ gradients outside that region. 

To test for systematics in the modelling, we explore the two extreme assumptions for the shape of the velocity ellipsoid: we use either JAM$_\text{cyl}$, with a cylindrically-aligned velocity ellipsoid \citep{Cappellari2008}, or JAM$_\text{sph}$, with a spherically-aligned orientation \citep{Cappellari_2020}. The anisotropy has a different meaning in the two cases. For JAM$_\text{cyl}$, the anisotropy is $\beta_z\equiv1-\sigma_z^2/\sigma_R^2$, where $\sigma_z$ and $\sigma_R$ are the intrinsic stellar velocity dispersions in the $R$ and $z$ cylindrical coordinates, respectively. For JAM$_\text{sph}$, the anisotropy is $\beta_r\equiv1-\sigma_\theta^2/\sigma_r^2$, where $\sigma_\theta$ and $\sigma_r$ are the intrinsic stellar velocity dispersions in the $r$ and $\theta$ spherical coordinates, respectively.

The JAM models have three free parameters (in addition to the usual nuisance parameters): the SMBH mass, the total mass-to-light ratio $(M/L)_\text{tot}$ and the orbital anisotropy $\beta_z$ or $\beta_r$. As standard with JAM, we parameterise the inclination with the intrinsic axis ratio $q_{\rm min}$ of the flattest Gaussian of the MGE. This is related to the galaxy inclination via Eq.~(14) of \citet{Cappellari2008}, where $q'_{\rm min}$ is the observed (i.e.\ projected) axis ratio of the flattest Gaussian of the MGE (see Table~\ref{tab:MGE components}):
\begin{equation}
i=\arctan\left(\sqrt{\frac{1-(q'_{\rm min})^2}{(q'_{\rm min})^2-q_{\rm min}^2}}\,\right)\,\,\,.
\end{equation}

A key practical difference between the Schwarzschild orbit-superposition method \citep{Schwarzschild1979} and JAM concerns the spatial extent of the kinematic data required. Schwarzschild modelling generally necessitates large-scale stellar kinematics, extending significantly beyond the immediate vicinity of a SMBH, to robustly constrain the orbital distribution and, consequently, the SMBH mass. In contrast, JAM typically yields the most reliable results when the kinematic fit is confined to a relatively small field of view (FoV). This FoV needs to be just large enough to effectively break the degeneracy between $M_{\mathrm{BH}}$ and the total mass-to-light ratio, $(M/L)_{\mathrm{tot}}$.

The underlying reason for this difference lies in the intrinsic flexibility of the models. A Schwarzschild model possesses significant freedom when constructing the orbital distribution; without the constraints provided by extensive kinematic data, the orbital structure remains poorly determined, diminishing the model's predictive power regarding $M_{\mathrm{BH}}$. Consequently, detailed modelling incorporating variations in the dark matter halo profile and stellar $M/L$ gradients across the galaxy is often essential for accurate results with this method. Conversely, JAM operates under stronger assumptions, where the stellar kinematics is uniquely determined by a few global parameters (e.g.\ describing the anisotropy profile and the $M/L$). This reduced flexibility means JAM does not inherently require large-scale data to constrain its parameter space. As a result, within the limited spatial region typically used for JAM fits, one can often approximate both the stellar $M/L$ and the velocity anisotropy as constant, simplifying the modelling compared to a full Schwarzschild approach. Furthermore, the stellar $M/L$ gradient is constrained by our CO dynamical modelling to be small, and therefore the assumption of a constant $M/L$ over the small FoV required for JAM modelling is reasonable.

The optimal spatial region for a JAM fit should span from within the SMBH's sphere of influence ($R_{\rm SoI}$) to a radius several times, but generally not more than an order of magnitude, larger than $R_{\rm SoI}$. This ensures the inclusion of kinematic data from the region where the SMBH's gravitational potential dominates stellar motions, as well as from a region where its influence is negligible, allowing the separation of the SMBH mass signature from the stellar mass contribution.

This distinction in methodology and data requirements was explored in detail by \citet{Thater2022}. They demonstrated that the $M_{\mathrm{BH}}$ recovered using JAM, by fitting kinematics within only the central $5$~arcsec in radius of their target while assuming a constant $M/L$ and neglecting a putative dark matter halo, were consistent with results from comprehensive Schwarzschild models fitted out to a radius of $30$~arcsec that explicitly included a dark matter component. \citet{Thater2022} showed this agreement arises because the total dynamical $M/L$ derived from JAM within the central region matched the total $M/L$ (including stars and dark matter) derived from the Schwarzschild model within that same central region, even though the Schwarzschild model's $M/L$ increased at larger radii due to the contribution of the dark halo \citep[see fig.~D1 in][]{Thater2022}.

We perform the fits and derive the corresponding formal uncertainties using the \textsc{CapFit} least-squares fitting procedure\footnote{\url{https://pypi.org/project/ppxf/}} (see Sec.~4.2 of \citealt{Cappellari2023}). We start by assigning constant uncertainties $\Delta V_\text{rms}=1$~km~s$^{-1}$ to all kinematic measurements and perform the fit with \textsc{CapFit}. This yields the best-fitting parameters $p_j$ and their formal uncertainties $\Delta p_j$ (from the diagonal terms of the covariance matrix at the best fit) at the $1\sigma$ confidence level. We then make the common assumption of a formally good fit (as indicated by our data-model comparison) to rescale the uncertainties in such a way as to obtain $\chi^2_\text{red} \equiv \chi^2/\mathrm{DOF}=1$, where DOF is the number of degrees of freedom. This is achieved by setting the new formal uncertainties $\Delta p_j\leftarrow\Delta p_j\sqrt{\chi^2_\text{red}}$. 

Contrary to the case of our CO SMBH determinations in Section~\ref{section:Dynamical Modelling}, here we did not apply the approximate $(2N)^{0.25}$ scaling of the uncertainties of \citet{Mitzkus2017}, because the crude correction for very large datasets is no more justified. In fact, we are fitting just $35$ $V_{\rm rms}$ data points and standard statistics can be used. However, this implies that the uncertainties of the two SMBH determinations are not directly comparable.

\begin{figure}
	\centering 
    \vspace{0.2cm}
	\includegraphics[width=\columnwidth]{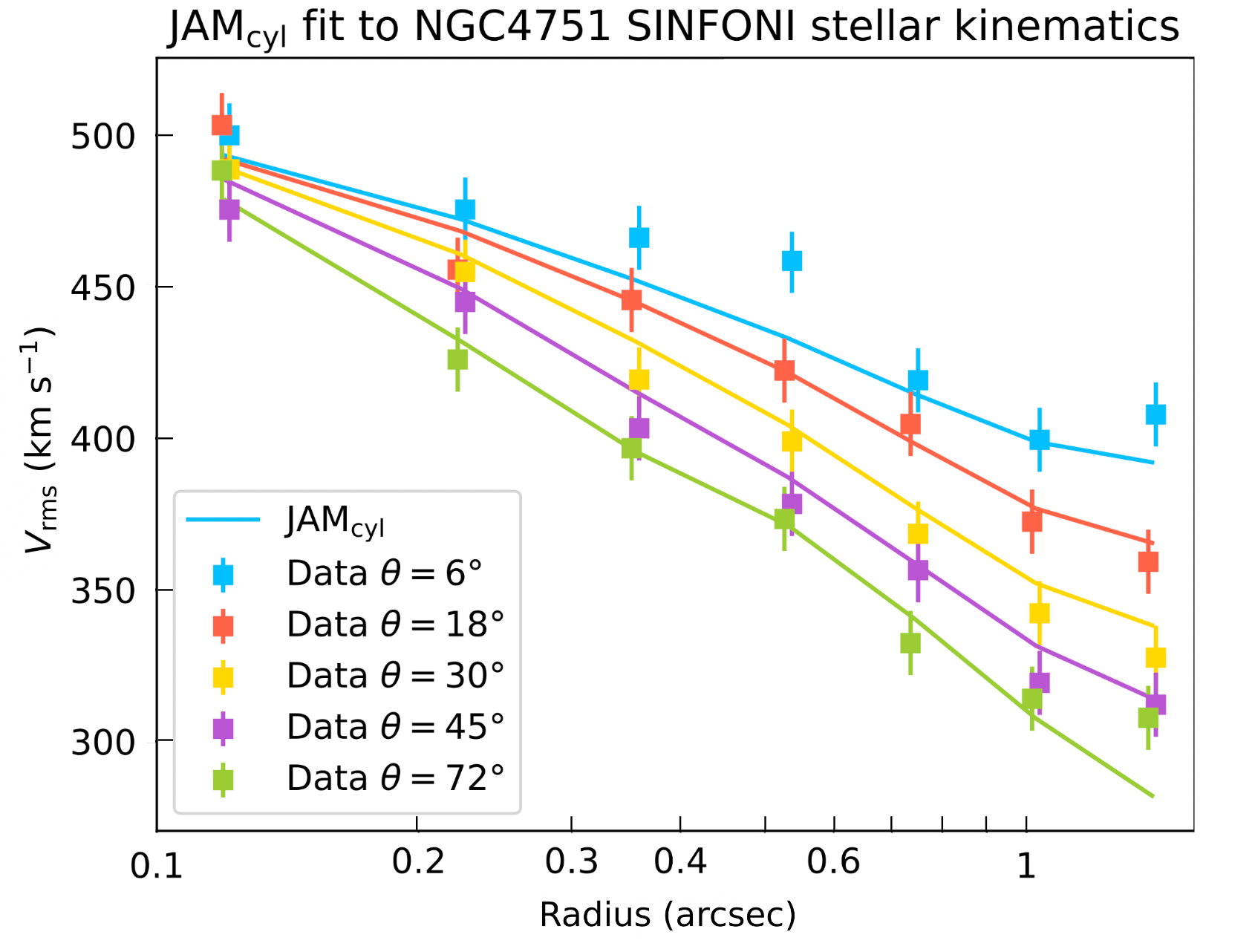}
    \newline
    \includegraphics[width=1.02\columnwidth]{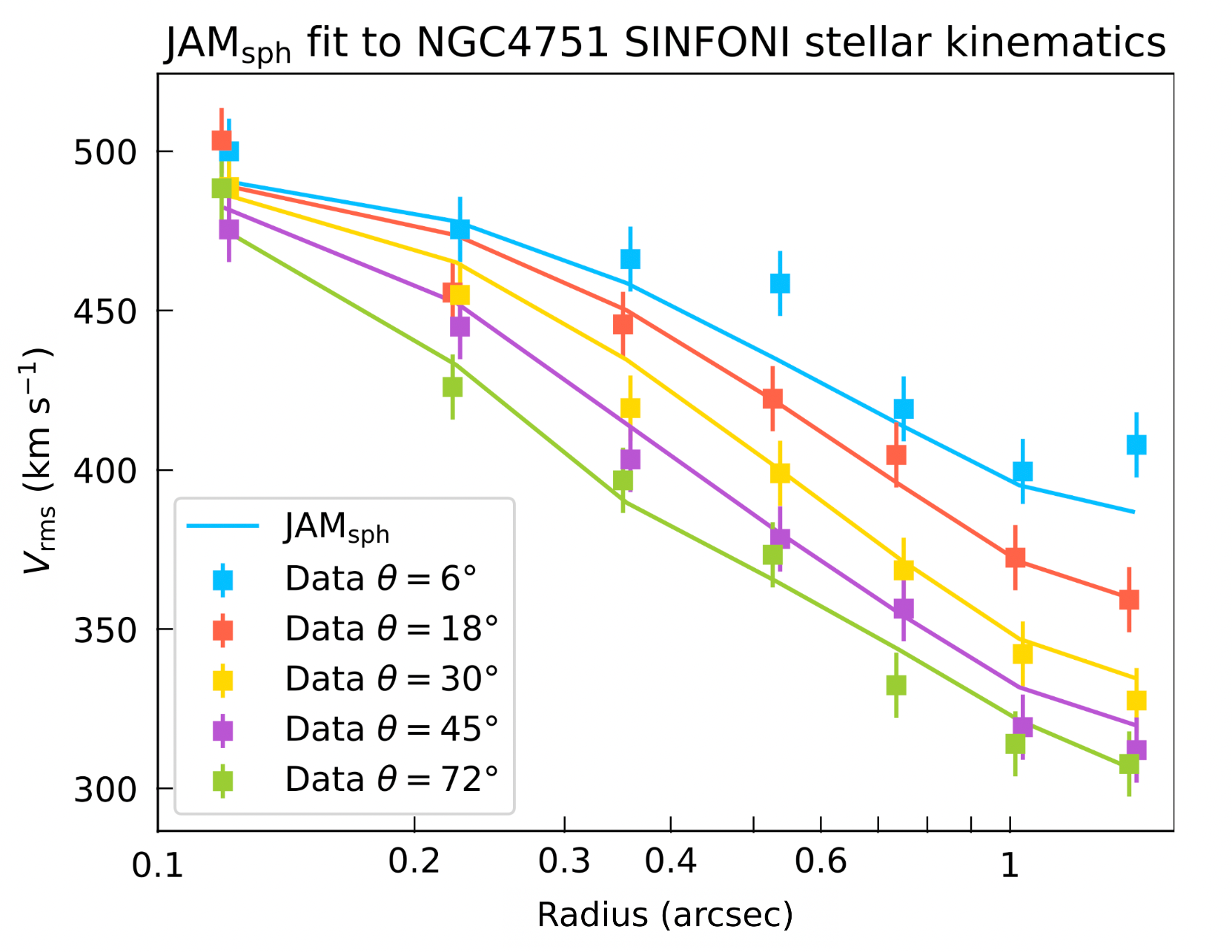}
	\caption{JAM$_\text{cyl}$ (top) and JAM$_\text{sph}$ (bottom) modelling of the SINFONI stellar kinematics from \citet{Rusli2013}.
    The data are shown for five angular sectors (as indicated) centred on the galaxy nucleus. The uncertainties are scaled to yield $\chi^2_\text{red}=1$.}
    \label{fig:jam_bh}
\end{figure}

The results of the JAM fits are shown in Fig.~\ref{fig:jam_bh}. The data-model comparisons are shown for the five angular sectors provided by \citet{Rusli2013}. The models fit the data to a high accuracy. To obtain $\chi^2_\text{red}=1$, one must set $\Delta V_\text{rms}=10$~km~s$^{-1}$, corresponding to a median random uncertainty of the stellar kinematics of just $2.6$~per~cent, at the expected level for good-quality SINFONI data. The best-fitting parameters and their $1\sigma$ formal uncertainties are $M_\text{BH}=(2.52\pm 0.12)\times10^9$~M$_\odot$ and $(M/L)_\text{tot}=3.759\pm0.094$~M$_\odot$/L$_{\odot,\text{F160W}}$ for JAM$_\text{cyl}$, and $M_\text{BH}=(3.24\pm0.29)\times10^9$~M$_\odot$ and $(M/L)_\text{tot}=4.22\pm0.16$~M$_\odot$/L$_{\odot,\text{F160W}}$ for JAM$_\text{sph}$. These best-fitting parameters are remarkably consistent with those obtained from our molecular gas modelling (see Section~\ref{section:Results}) and are summarised in Table~\ref{tab:stellar kinematics}. The best-fitting inclination for both JAM models is $i\approx72\degree$, also close to that derived from the molecular gas kinematics.

\begin{table}
    \begin{threeparttable}
    \renewcommand\thetable{5}
    \caption{Best-fitting stellar kinematic model parameters and associated uncertainties.}
    \centering
    \begin{tabular}{lccc}
        \hline
         Parameter & Best fit & $1\sigma$ uncertainty & $3\sigma$ uncertainty \\ 
         \hline
        \multicolumn{4}{c}{\textbf{JAM$_\text{cyl}$ model}}\\
        \hline
         $M_\text{BH}$~($10^9$~M$_\odot$) & $\phantom{-}2.52$ & $\pm0.12$ & $\pm0.36$\\
         $(M/L)_{\text{tot}}$ (M$_\odot$/L$_{\odot,\text{F160W}}$) & $\phantom{-}3.76$ & $\pm0.09$ & $\pm0.28$\\ 
         $\beta_\text{z}$ & $-0.23$ & $\pm0.09$ & $\pm0.26$\\
         $q_\text{min}$ & $\phantom{-}0.05$ & $-0.05, +0.08$ & $-0.05, +0.24$\\
         \hline
        \multicolumn{4}{c}{\textbf{JAM$_\text{sph}$ model}}\\
        \hline
        $M_\text{BH}$~($10^9$~M$_\odot$) & $\phantom{-}3.24$ & $\pm0.29$ & $\pm0.87$\\
         $(M/L)_{\text{tot}}$ (M$_\odot$/L$_{\odot,\text{F160W}}$) & $\phantom{-}4.22$ & $\pm0.16$ & $\pm0.48$\\
         $\beta_\text{r}$ & $-3.1\phantom{0}$ & $\pm2.1\phantom{0}$ & $\pm6.3\phantom{o}$\\
         $q_\text{min}$ & $\phantom{-}0.05$ & $-0.05, +0.07$ & $-0.05, +0.22$\\
         \hline
    \end{tabular}
    \label{tab:stellar kinematics}
    \end{threeparttable}
\end{table}

Galaxies are generally expected to be close to isotropic ($\beta_z=0$ and $\beta_r=0$), with a tendency for the anisotropy to be tangential ($\beta_z<0$ and $\beta_r<0$) close to the SMBH (see Fig.~2 of \citealt{Cappellari_2008_inproceedings}). Here we are fitting the SINFONI data near the SMBH and therefore expect the anisotropy to be tangential, as observed. However, the anisotropy is poorly constrained, and for both the cylindrical and spherical alignment of the velocity ellipsoid, an isotropic model is also consistent with the data at the $3\sigma$ confidence level.


\section{Discussion}
\label{section:Discussion}

\subsection{Best-fitting molecular gas mass models}
\label{subsection: best-fitting CO mass models}

In this paper we presented molecular gas-based mass models with two different $M/L$ radial profiles. As shown in Figs.~\ref{fig:PVD_NGC4751} and \ref{fig:PVD_NGC4751_linearML}, the two models provide near-identical fits visually. Whilst the $\chi^2$ statistic is primarily used to compare models against data, in a limited number of situations it can also be used to compare model against model. This is the case here, as the model with a constant $M/L$ is nested within the model with a linearly-varying $M/L$ (i.e.\ the model with a constant $M/L$ is a particular case of the model with a linearly-varying $M/L$,  where the $M/L$ gradient is equal to $0$). We can thus judge the relative goodness-of-fit of the two models more robustly by using the reduced $\chi^2$ statistic $\chi^2_\text{red}$. The best-fitting CO model with a constant $M/L$ has $\chi^2_\text{red}=2.49$, while the best-fitting CO model with a radially linearly-varying $M/L$ has $\chi^2_\text{red}=2.43$. As such, we can conclude that the model with a linearly-varying $M/L$ is a marginally better fit (the additional $M/L$ gradient variable would be expected to converge to $0$ if the data strongly suggested a constant $M/L$). As the $\chi^2_\text{red}$ difference is small, however, we choose to also report the Bayesian information criterion $\text{BIC}\equiv k\ln N-2\ln P$ to compare the models, where $k$ is the number of free parameters of each model and $N$ and $\ln P$ are as defined in Section~\ref{section: Modelling process}. The BIC provides a heuristic approximation of the Bayes factor, which is the statistically-correct way to compare models. Indeed, the BIC not only allows for the comparison of non-nested models with different numbers of parameters, but it also penalises models with excessive dimensionality that could otherwise overfit the data. The BIC difference between the two best-fitting CO models presented in this paper is $\Delta_\text{BIC}\approx1$, and hence statistically-speaking it is not possible to formally prefer one model over the other. These results are summarised in Table~\ref{tab:chi2}.

\begin{table}
    \renewcommand\thetable{6}
	\centering
	\caption{Statistics of the models discussed.}
	\label{tab:chi2}
	\begin{tabular}{lccc} 
	\hline
	Model & $\chi^2_\text{red}$ & $k$ & BIC\\
        (1) & (2) & (3) & (4)\\
	\hline
        Best-fitting CO model with constant $M/L$ & $2.49$ & $10$ & $\phantom{1}610$\\
        Best-fitting CO model with linearly-varying $M/L$ & $2.43$ & $11$ & $\phantom{1}609$\\
	\hline
    \end{tabular}
    \begin{tablenotes}
          \item \textit{Notes.} (1) Model. (2) Reduced $\chi^2$ statistic. (3) Number of free parameters.\\ (4) Bayesian information criterion.
    \end{tablenotes}
\end{table}

\begin{figure}
    \centering
    \includegraphics[scale=0.55]{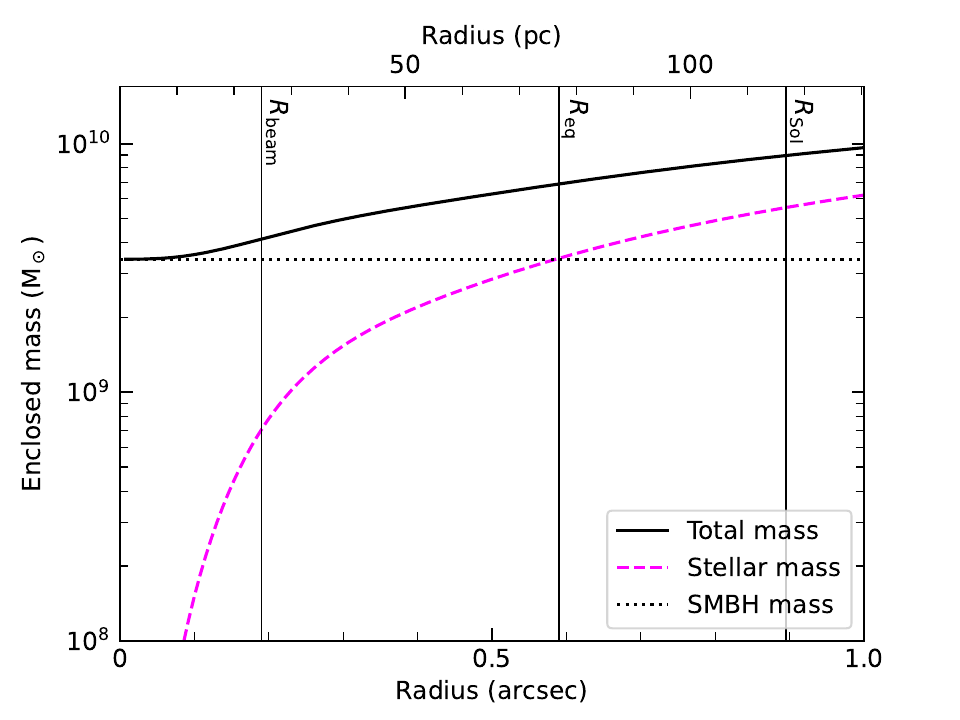}
    \caption{Cumulative mass function of NGC~4751 for the molecular gas kinematic model with a constant $M/L$, showing the relative contributions of the SMBH (black dotted line) and stars (magenta dashed line) to the total enclosed mass (black solid line). The three vertical black lines indicate the physical extent of the synthesised beam ($R_\text{beam}$), the radius of the SMBH sphere of influence ($R_{\text{SoI}}$, assuming $\sigma_\text{e}=355.4$~km~s$^{-1}$ and our best-fitting SMBH mass) and the radius of equal mass contribution ($R_{\text{eq}}$).}
    \label{fig:cummulative mass function}
\end{figure}

\begin{figure}
    \centering
    \includegraphics[scale=0.55]{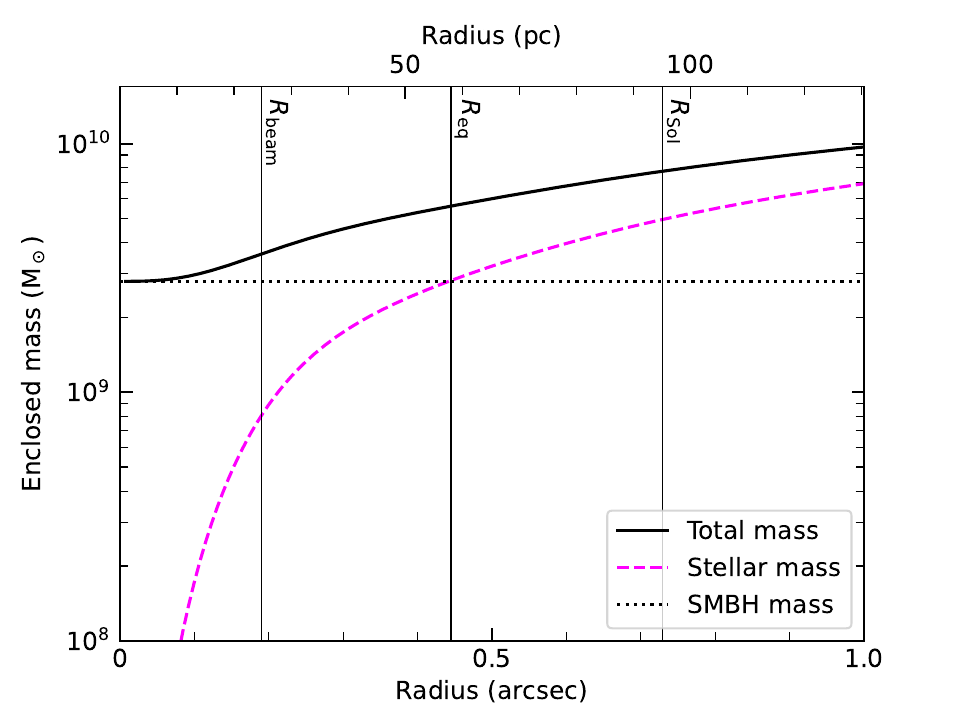}
    \caption{As Fig.~\ref{fig:cummulative mass function}, but for the model with a radially linearly-varying $M/L$.}
    \label{fig:cummulative mass function linear}
\end{figure}

Figures~\ref{fig:cummulative mass function} and \ref{fig:cummulative mass function linear} show the cumulative mass distribution of NGC~4751 for the molecular gas kinematic model with a constant $M/L$ and the model with a radially linearly-varying $M/L$, respectively. Using the standard definition of a SMBH SoI (see Section~\ref{section:NGC4751}) and the best-fitting SMBH mass for the constant $M/L$ model yields $R_{\text{SoI}}\approx116$~pc ($\approx0.89$~arcsec), whereas the best-fitting SMBH mass for the linear $M/L$ model yields $R_{\text{SoI}}\approx95$~pc ($\approx0.73$~arcsec). Both of these are significantly larger than the size of the synthesised beam  ($R_\text{beam}$). One can also assess the impact of a SMBH by considering the radius at which the enclosed stellar mass is equal to that of the SMBH. Figure~\ref{fig:cummulative mass function} shows this radius of equal mass contribution to be $R_{\text{eq}}\approx77$~pc ($\approx0.59$~arcsec), smaller than the usual $R_{\text{SoI}}$ but still significantly larger than $R_\text{beam}$. Figure~\ref{fig:cummulative mass function linear} shows the radius of equal mass contribution to be $R_{\text{eq}}\approx59$~pc ($\approx0.45$~arcsec), also smaller than the usual $R_{\text{SoI}}$, but still more than twice $R_\text{beam}$. In fact, in both models the SMBH dominates the potential so significantly that at the innermost radius probed ($R_\text{beam}$) $\approx82$~per~cent of the enclosed mass is due to the SMBH in the constant $M/L$ model and $\approx 76$~per~cent in the linear $M/L$ model, making the SMBH mass determination trivial in both models.

The quality of a SMBH measurement can also be assessed in terms of how well resolved the SMBH SoI is and the proximity of the dynamical tracer to the SMBH. In the region dominated by the SMBH, \citet{Zhang_2024} derived a simple relation between the highest circular velocity measured $v_\text{c}$ and the innermost radius probed $R_\text{min}$, where $R_\text{min}$ can be normalised by respectively the Schwarzschild radius ($R_\text{Schw}\equiv2GM_\text{BH}/c^2$, where $c$ is the speed of light), $R_\text{SoI}$ and $R_\text{eq}$. The resulting relations allow to compare the spatial resolutions of different datasets and the sizes of the regions probed in a $M_\text{BH}$-independent manner. Our observations of NGC~4751 detect a maximum line-of-sight velocity $v_\text{obs}\approx687$~km~s$^{-1}$ (see e.g.\ Fig.~\ref{fig:PVD_NGC4751}). Deprojecting this velocity using $v_\text{c}=v_\text{obs}/\sin i$, where $i=78\fdg7$ (Table~\ref{tab:Best Fit}), the highest circular velocity probed is $\approx700$~km~s$^{-1}$. This is the second highest circular velocity ever probed by a molecular gas measurement, rivaled only by the high-resolution measurement of NGC~0383 \citep{Zhang_2025}. While some of the high central velocities must originate from the dense central stellar component of the galaxy (see Fig.~\ref{fig:HST}), our molecular gas data with a spatial resolution of $\approx24$~pc ($0.19$~arcsec) allow us to probe remarkably close to the SMBH: 
down to $R_\text{min}/R_\text{Schw}\approx83,000$, $R_\text{min}/R_\text{SoI}\approx0.23$ and $R_\text{min}/R_\text{eq}\approx0.33$ for the model with a constant $M/L$, and $R_\text{min}/R_\text{Schw}\approx101,000$, $R_\text{min}/R_\text{SoI}\approx0.28$ and $R_\text{min}/R_\text{eq}\approx0.46$ for the model with a radially linearly-varying $M/L$.


\subsection{Molecular gas-derived SMBH mass uncertainties}
\label{subsection:uncertainties}

Aside from the fitting uncertainties listed in Table~\ref{tab:Best Fit}, which are fundamentally statistical in nature, our molecular gas measurements are subject to several potential sources of systematic error. The first of these is the distance. For all dynamical mass measurements, $M_\text{BH}\propto D$, so the adopted distance affects the SMBH mass linearly. Conversely, the inferred SMBH mass can be re-scaled for any adopted distance. As standard, the distance uncertainties are neither quoted nor propagated into our SMBH mass measurement, but these are typically about $20$ -- $30$~per~cent when flow models are applied to relatively nearby galaxies ($D<50$~Mpc; \citealt{Lelli_2016}). 

In our modelling we have only considered the mass contributions of the SMBH and stars, and have assumed the dark matter and molecular gas to have negligible masses. The spatial scales over which dark matter effects are significant are much greater than those considered in this paper and those where the SMBH dominates the potential. In the very unlikely scenario that there is significant diffuse dark matter in the central regions, its contribution to the circular velocity is expected to increase with radius similarly to that of the stars, whereas the contribution of the SMBH declines in a Keplerian fashion (see e.g.\ Fig.~7 of \citealt{Lelli_2022}). Thus, the dark matter contribution is most strongly degenerate with that of the stars, which would lead to a different best-fitting $M/L$ without strongly affecting the SMBH mass. Furthermore, \citet{Rusli2013} considered models of NGC~4751 with and without dark matter, and there was no significant difference in the SMBH masses inferred, with only a minor impact on the $M/L$ (see Section~\ref{section:NGC4751}). Similarly, as discussed in Section~\ref{section:line emission}, NGC~4751 has an extended molecular gas distribution with a total mass of $(2.27\pm0.01)\times10^8$~M$_\odot$, which if plotted would not even be visible in Fig.~\ref{fig:cummulative mass function} (as the total mass is only reached at the outer edge of the disc). The contribution of molecular gas is thus negligible compared to those of the SMBH and stars. 

Another source of uncertainties of the model parameters reported in this work comes from the modelling assumptions made. Our modelling involves some trade-offs between allowing more model complexity, which should lead to more accurately reproduce the data, and introducing simplifying assumptions, to ensure model convergence and a reasonable computational time. Our models clearly include many simplifying assumptions, such as a constant position angle and inclination angle, but these assumptions are well-substantiated by the data themselves (see Fig.~\ref{fig:mom012}). Our exploration of different $M/L$ radial profiles appears in this case to be a useful test of how variations of an important model assumption affect the model outcomes, and the differences between the best-fitting $M_\text{BH}$ and $M/L$ of the two models are a reflection of systematic errors.

In addition to the two models presented in this paper, we tested alternative models to better quantify potential systematic errors. In the previous sections, we used an exponential surface brightness profile to parameterise the molecular gas distribution. An alternative and more appropriate, but often more computationally costly, way of representing the molecular gas distribution is to simply use the molecular gas distribution itself as a model input. This can be done through the use of either the zeroth-moment map or the clean components generated during the cleaning process. We chose to create such a model using the latter method and \textsc{SkySampler}\footnote{\url{https://github.com/Mark-D-Smith/KinMS-skySampler}} \citep{WISDOM_IV}, a software compatible with the \textsc{KinMS} environment. \textsc{SkySampler} generates a large number of particles whose densities and positions are determined by the clean components, and which are then processed by \textsc{KinMS} as before: they are deprojected (according to a position angle and inclination under the assumption of an infinitely-thin disc) and assigned line-of-sight velocities to create a model data cube. This yields a best-fitting SMBH mass $M_\text{BH}=2.94^{+0.36}_{-0.35}\times10^9$~M$_\odot$ and a best-fitting F160W filter $M/L_{F160W}=(2.78\pm0.11)$~M$_\odot$/L$_{\odot,\text{F160W}}$, representing a $17$~per~cent decrease of the SMBH mass and a $4$~per~cent increase of the $M/L$ compared to the exponential disc model with a constant $M/L$.

A potential source of systematic errors arises from the choice of MGE model. In this paper we use the the \textsc{MgeFit} procedure to parameterise the stellar light distribution, but this is not the only method available to us. \citet{Davidson_2024} present an alternative MGE model based on a 2D parametric galaxy fit obtained using the algorithm \textsc{GALFIT}. In addition to using a different algorithm, \citet{Davidson_2024} utilise a different PSF. These differences can have a significant impact on the innermost Gaussian of the MGE, particularly in cuspy galaxies like NGC~4751, which can in turn impact modelling outcomes. To test these effects, we performed a fit using the \citet{Davidson_2024} MGE (their Table~5; with uniform PA), yielding a best-fitting SMBH mass $M_\text{BH}=2.79^{+0.59}_{-0.61}\times10^9$~M$_\odot$ and a best-fitting F160W filter $M/L_{F160W}=2.42_{-0.11}^{+0.13}$~M$_\odot$/L$_{\odot,\text{F160W}}$, representing a $19$~per~cent decrease of the SMBH mass and a $10$~per~cent decrease of the $M/L$ compared to those of the model with the MGE components from Table~\ref{tab:MGE components}.

Another likely major source of systematic errors arises from noticeable dust extinction in the NGC~4751 \textit{HST} image (Fig.~\ref{fig:HST}). While care was taken to choose an image in the reddest filter possible (to minimise dust extinction) and we masked the most obviously dust-obscured pixels, a more thorough treatment of dust involves creating a dust-corrected MGE model. We now attempt this below.

We perform the dust correction using a WFC3 \textit{F}110\textit{W} filter ($J$-band) image in addition to the previously used F160W filter ($H$-band) image. We first re-drizzle the \textit{F}110\textit{W} filter image to match the pixels in the F160W filter image. Having converted both images to units of magnitude in the Vega system, we calculate the observed $J-H$ colour of every pixel $(J-H)_\text{obs}$. Given that the inner region of the galaxy is reddened by the embedded dust disc, we can only confidently estimate the intrinsic $J-H$ colour on the outskirts of the image, beyond the large-scale dusty circumnuclear disc. We measure $(J-H)_\text{int}=0.8$~mag and assume this number throughout. We then calculate the colour excess of every pixel $E(J-H)\equiv(J-H)_\text{obs}-(J-H)_\text{int}$ and create an extinction map $A_H$ by dividing the colour excess map by the ratio of the total extinctions obtained using the standard Galactic extinction curve with $R_V=3.1$ \citep{Fitzpatrick_1999}, where $R_V$ is the total-to-selective extinction ratio:
\begin{equation}
    A_H=\frac{E(J-H)}{\bigg(\cfrac{A_J}{A_V}\cfrac{A_V}{A_H}-1\bigg)}\,\,\,. 
\end{equation}
\noindent We then correct our F160W filter image in the usual manner using this extinction map. Inspection of this extinction-corrected F160W filter image shows that whilst the effects of dust have been significantly alleviated, the most prominent dust lane close to the central region of the galaxy remains, suggesting that the dust in that region is (close to being) optically thick. We thus proceed to create two different MGE models, based on how we choose to treat this remaining dust.

\begin{figure}
    \centering
    \includegraphics[scale = 0.7, trim = {0 0 1cm 0}]{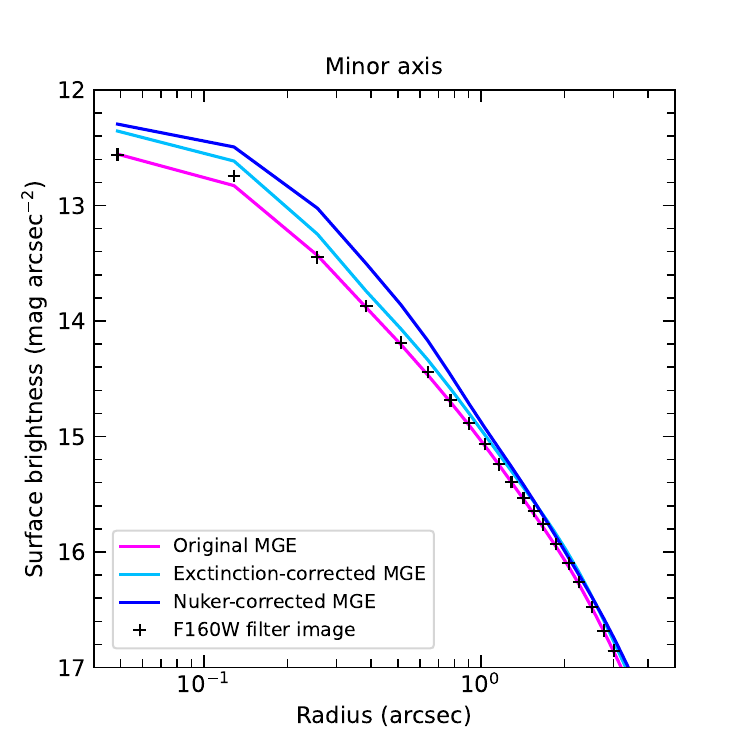}
    \caption{Central surface brightness profiles as a function of elliptical radius along the far side of the minor axis. Included are the profiles of the F160W~filter image (black crosses), the original MGE (magenta line), the extinction-corrected MGE (cyan line) and the Nuker-corrected MGE (blue line).
    }
    \label{fig:MGE profiles}
\end{figure}

In the first case, we simply use the same hand-drawn mask used in our original MGE model (see Fig.~\ref{fig:MGE contours}), to exclude the near side of the dusty disc that most affects the galaxy surface brightness. We then create a new MGE (using the procedure outlined in Section~\ref{section: Modelling process}), which we refer to as the "extinction-corrected MGE", whose "deconvolved" Gaussian components are listed in Table~\ref{tab:Alternative MGE}. The minor-axis surface brightness profile of this extinction-corrected MGE is shown in Fig.~\ref{fig:MGE profiles} alongside that of the original MGE. This demonstrates that our dust correction strongly enhanced the central regions of the galaxy, especially the maximum (i.e.\ central) surface brightness. This extinction-corrected MGE model leads to a best-fitting SMBH mass $M_\text{BH}=3.21^{+0.49}_{-0.43}\times 10^9$~M$_\odot$ and a best-fitting F160W filter $M/L_{F160W}=(2.42\pm0.10)$~M$_\odot$/L$_{\odot,\text{F160W}}$, representing a $7$~per~cent decrease of the SMBH mass and a $11$~per~cent decrease of the $M/L$ compared to our exponential disc model with a constant $M/L$.

 \begin{table}
    \renewcommand\thetable{7}
	\centering
	\caption{Parameters of the "deconvolved" best-fitting dust-corrected MGE components.}
	\label{tab:Alternative MGE}
	\begin{tabular}{c@{\hskip5pt}c@{\hskip8pt}c|c@{\hskip5pt}c@{\hskip8pt}c} 
    \multicolumn{3}{c}{Extinction-corrected} & \multicolumn{3}{c}{Nuker-corrected}\\
	\hline
	$\log I_{\odot}^\prime$ & $\log \sigma^\prime$ & $q^\prime$ & $\log I_{\odot}^\prime$ & $\log \sigma^\prime$ & $q^\prime$ \\
        (L$_{\odot, \text{F160W}}$~pc$^{−2}$) & (arcsec) & &(L$_{\odot, \text{F160W}}$~pc$^{−2}$) & (arcsec) & \\
        (1) & (2) & (3) &  (1) & (2) & (3) \\
	\hline
	5.488 & $-0.935$ & 0.951 & 5.577 & $-0.922$ & 0.990 \\
	5.007 & $-0.590$ & 0.405 & 5.028 & $-0.363$ & 0.623\\
	4.771 & $-0.446$ & 0.790 & 4.183 & $-0.140$ & 0.841\\
    4.506 & $-0.272$ & 0.990 & 4.518 & $\phantom{-}0.043$ & 0.351\\
    4.033 & $\phantom{-}0.088$ & 0.990 & 4.212 & $\phantom{-}0.235$ & 0.708\\
    4.255 & $\phantom{-}0.103$ & 0.327 & 3.991 & $\phantom{-}0.466$ & 0.556\\
    3.928 & $\phantom{-}0.328$ & 0.441 & 3.631 & $\phantom{-}0.787$ & 0.456\\
    4.017 & $\phantom{-}0.534$ & 0.512 & 2.702 & $\phantom{-}1.255$ & 0.334\\
    3.526 & $\phantom{-}0.935$ & 0.401 & 3.029 & $\phantom{-}1.255$ & 0.778\\
    2.923 & $\phantom{-}1.412$ & 0.416 & \multicolumn{3}{c}{$-$}\\
	\hline
    \end{tabular}
    \begin{tablenotes}
          \item \textit{Notes.} "Deconvolved" MGE Gaussian components. (1) Central surface brightness. (2) Standard deviation. (3) Axial ratio.
    \end{tablenotes}
\end{table}

In the second case, given that the standard extinction correction is insufficient to correct the dustiest portions of the image, we perform an additional correction in the following manner. Firstly, we identify the dustiest pixels of our original $F160W$ filter image by creating a "dust mask", selecting all pixels with $(J-H)_\text{obs}>0.9$~mag. Having determined the outer position angle and axis ratio of the galaxy, we calculate the elliptical radius of every pixel in the dust-masked image to obtain a radial surface brightness profile which we fit with a 1D Nuker model \citep{Lauer_1995}. The Nuker model is a double-power law with a break radius $r_\text{b}$, that represents the transition between the inner and outer slopes $\gamma$ and $\beta$, a surface brightness at the transition break radius $I_\text{b}$ and a transition sharpness $\alpha$. We perform the fit to the data using the \texttt{scipy.optimize.curve\_fit} routine and obtain $r_\text{b}=0.45$~arscec, $I_\text{b}=13.3$ mag~arcsec$^{-2}$, $\alpha=1.48$, $\beta=1.60$ and $\gamma=0.00$. We note that this and subsequent fits represent the observed, and not intrinsic, Nuker parameters. We then repeat these dust-masking and profile-fitting steps with the extinction-corrected $F160W$ filter image used for the previous fit, whilst keeping the break radius fixed to $r_\text{b}=0.45$~arcsec. This yields $I_\text{b}=13.2$ mag~arcsec$^{-2}$, $\alpha=1.26$, $\beta=1.67$ and $\gamma=0.01$. Finally, we replace the surface brightnesses of the dust-masked pixels in the extinction-corrected image with those predicted by the best-fitting Nuker model, and proceed with a MGE as before, which we refer to as the "Nuker-corrected MGE", whose "deconvolved" Gaussian components are listed in Table~\ref{tab:Alternative MGE}. As before, the minor-axis surface brightness profile is shown in Fig.~\ref{fig:MGE profiles} alongside the other models. This Nuker-corrected MGE leads to a best-fitting SMBH mass $M_\text{BH}=3.65^{+0.39}_{-0.37}\times 10^9$~M$_\odot$ and a best-fitting F160W filter $M/L_{F160W}=1.88^{+0.08}_{-0.07}$~M$_\odot$/L$_{\odot,\text{F160W}}$, representing a $6$~per~cent increase of the SMBH mass and a $30$~per~cent decrease of the $M/L$ compared to the exponential disc model with a constant $M/L$.

It is worth highlighting that there are other, more involved dust-correction methods that we could implement, such as an embedded screen dust model. However, following the embedded screen model of \citet{Viaene_2017} and \citet{Boizelle2019} results in less than $10$~per~cent higher dust correction along the major axis, in large part due to our use of a near-infrared image. As this does not change the best-fitting dynamical modelling parameters significantly more than the spread of the two MGE models above, we do not attempt this more detailed dust correction here.

For all four alternative models, we also carried out analogous tests using a radially linearly-varying $M/L$. The results of all the tests are listed in Table~\ref{tab:systematic error}.

\begin{table*}
    \begin{threeparttable}
    \renewcommand\thetable{8}
    \caption{Best-fitting molecular gas model parameters and associated uncertainties, for the models explored in Section~\ref{subsection:uncertainties}.}
    \centering
    \begin{tabular}{lc@{\hskip4pt}rc@{\hskip4pt}rc@{\hskip4pt}rc@{\hskip4pt}rc@{\hskip4pt}r}
        \hline
        Parameter & \multicolumn{2}{c}{Original model} & \multicolumn{2}{c}{\textsc{SkySampler}} & \multicolumn{2}{c}{\citet{Davidson_2024}
        } & \multicolumn{2}{c}{Extinction-corrected} & \multicolumn{2}{c}{Nuker-corrected}\\
        & Best fit & $3\sigma$ & Best fit & $3\sigma$ & Best fit & $3\sigma$ & Best fit & $3\sigma$ & Best fit & $3\sigma$ \\
        \hline
        \multicolumn{11}{c}{\textbf{Models with a constant $M/L$}}\\
         \hline        $\log(M_\text{BH}/\text{M}_\odot)$ & $9.54$ & $-0.06, +0.05$ & $9.47$ & $-0.06,+0.05$ & $9.45$ & $-0.11, +0.08$ & $9.51$ & $\pm0.06$ & $9.56$ & $-0.05, +0.04$\\        
         $M/L_{\text{F160W}}$ & $2.68$ & $\pm0.11$ & $2.78$ & $\pm0.11$ & $2.42$ & $-0.11, +0.13$ & $2.42$ & $\pm0.10$ & $1.88$ & $-0.07, +0.08$\\
        \hline
        \multicolumn{11}{c}{\textbf{Models with a radially linearly-varying $M/L$}}\\
        \hline        $\log(M_\text{BH}/\text{M}_\odot)$ & $9.45$ & $\pm0.10$ & $9.39$ & $-0.09, +0.10$ & $9.37$ & $-0.14, +0.13$ & $9.43$ & $-0.09, +0.08$ & $ 9.55$ & $-0.08, +0.06$\\        
        $M/L_{\text{F160W}}$ intercept & $3.07$ & $-0.35, +0.27$ & $3.15$ & $-0.38, +0.27$ & $2.60$ & $-0.27, +0.21$ & $2.71$ & $-0.22, +0.23$ & $1.93$ & $-0.21, +0.25$\\        
        $M/L_{\text{F160W}}$ gradient & $-0.09\phantom{-}$ & $-0.06, +0.08$ & $-0.10\phantom{-}$ & $-0.07, +0.10$ & $-0.04\phantom{-}$ & $-0.05, +0.07$ & $-0.07\phantom{-}$ & $-0.06, +0.05$ & $-0.01\phantom{-}$ & $-0.06, +0.05$\\
        \hline
    \end{tabular}
     \begin{tablenotes}
          \item \textit{Note.} The $M/L$ and $M/L$ intercept are expressed in units M$_\odot$/L$_{\odot,\text{F160W}}$ and the $M/L$ gradient is expressed in units M$_\odot$/L$_{\odot,\text{F160W}}$~arcsec$^{-1}$.
    \end{tablenotes}
    \label{tab:systematic error}
    \end{threeparttable}
\end{table*}

Considering all these tests, it is clear that the systematic variations of the SMBH mass are similar to the statistical uncertainties reported in Table~\ref{tab:Best Fit}, with $M_\text{BH}=3.43^{+0.45}_{-0.44}[\text{stat},\,3\sigma]^{+0.22}_{-0.64}[\text{sys}]\times10^9$~M$_\odot$ for the model with a constant $M/L$, where the first uncertainties are statistical and the second systematic. However, this is less clear when considering the $M/L$. Dust impacts the $M/L$ significantly, with $M/L_{F160W}=2.68\pm0.11[\text{stat},\,3\sigma]^{+0.10}_{-0.80}[\text{sys}]$~M$_\odot$/L$_{\odot,\text{F160W}}$ for the model with a constant $M/L$. For the radially linearly-varying $M/L$ case, the variations of the SMBH mass are also similar to the statistical uncertainties, with $M_\text{BH}=2.79^{+0.75}_{-0.57}[\text{stat},\,3\sigma]^{+0.75}_{-0.45}[\text{syst}]\times10^9$~M$_\odot$. And again dust has a significant impact on the $M/L$, with $\left(M/L_\text{F160W}\right)/\left(\text{M}_\odot/\text{L}_{\odot,\text{F160W}}\right)=3.07^{+0.27}_{-0.35}[\text{stat},\,3\sigma]^{+0.08}_{-1.14}[\text{sys}] -0.09^{+0.08}_{-0.06}[\text{stat},\,3\sigma]^{+0.08}_{-0.01}[\text{sys}]\,\left(R/\text{arcsec}\right)$. It is worth noting that the Nuker-corrected MGE model with a linearly-varying $M/L$ yields a gradient remarkably close to zero, suggesting that much of the $M/L$ gradient in previous models is due to the embedded dust disc.

These tests indicate that our method of rescaling the statistical uncertainties to approximate the systematic uncertainties is appropriate in certain contexts, but not in others. It is clear that the choice of MGE model has a significant impact on the inferred $M/L$ and that the dust extinction correction of optical images needs to be more systematically implemented in the field of molecular gas modelling. On the other hand, it is also clear that the inferred SMBH masses from this ALMA Cycle 4 dataset are largely unaffected by the MGE models and that even the simplest models yield robust SMBH masses. This is consistent with the general understanding that, as long as the kinematic tracers remain regular and probe deep within $R_\text{SoI}$, and the data resolve this $R_\text{SoI}$, the SMBH mass will be less susceptible to uncertainties in the galaxy's central stellar light distribution.


\subsection{Comparison of stellar kinematic results}
\label{Section: comparison}

Similarly to Section~\ref{subsection: best-fitting CO mass models}, we can calculate the SMBH SoI of our stellar kinematic best fits. The best-fitting mass for the model with JAM$_\text{cyl}$ yields $R_\text{SoI}\approx86$~pc ($\approx0.66$~arcsec), whereas that for the model with JAM$_\text{sph}$ yields $R_\text{SoI}\approx 110$~pc ($\approx0.85$~arcsec). As mentioned in Section~\ref{section:JAM models}, the field of view of the data modelled by JAM must be large enough to break the degeneracy between the SMBH mass and $(M/L)_\text{tot}$, but no more. Given that the SINFONI data probe radii up to $\approx1.6$~arcsec, they more-than-satisfactorily encompass the $R_\text{SoI}$ of both models, allowing for a break in the degeneracy.

As discussed in Section~\ref{section:NGC4751}, prior to our JAM models, \citet{Rusli2013} used the same SINFONI data to infer a SMBH mass of $(1.4\pm0.1)\times10^9$~M$_\odot$ at $1\sigma$ ($68$~per~cent) confidence level, modelled using three-integral Schwarzschild models and assuming the same distance as us. Unlike our JAM SMBH masses, the SMBH mass of \citet{Rusli2013} is not consistent with our molecular gas mass. It is not consistent with our JAM SMBH masses either.

Part of this disagreement may stem from a potential underestimation of uncertainties in the \citet{Rusli2013} determination. Examining their Fig.~11, it is clear that the $\Delta\chi^2$ line exhibits multiple minima. The primary minimum corresponds to the formal best-fitting solution, with a black hole mass of approximately $1.4 \times 10^9$~M$_\odot$, as reported in their paper. However, there is a secondary minimum at $1.8\times10^9$~M$_\odot$, lying outside the formal $3\sigma$ confidence interval. This secondary minimum likely arises from numerical noise, a common feature of the \citet{Schwarzschild1979} dynamical modelling method. It is also the reason why the associated $\Delta\chi^2$ are sometimes smoothed before determining confidence intervals (see e.g.\ \citealt{Gebhardt2003}). The sharp global minimum may also be an artifact of numerical noise. Arguably, a more realistic uncertainty would encompass the region where the $\Delta\chi^2$ profile sharply rises. This would correspond to a black hole mass range of approximately $[1.1, 2.0]\times10^9$~M$_\odot$ ($3\sigma$ limits). If this were the true confidence interval of the \citet{Rusli2013} models, it would significantly reduce, although not eliminate, the tension between their result and ours.

As discussed in Section~\ref{section:NGC4751}, the $M/L_R$ derived by \citet{Rusli2013} is unusually high, with $M/L_R=12.2^{+0.6}_{-0.7}$~M$_\odot/$~L$_{\odot,R}$ (based on a model with dark matter present) or $M/L_R=13.1^{+0.3}_{-0.4}$~M$_\odot/$~L$_{\odot,R}$ (based on a model without dark matter). Such an $M/L_R$ is inconsistent (approximately twice) that expected from similar targets \citep{Junqiang_2021}. \citet{Rusli2013} do not use an $R$-band image directly, but rather re-scale a F160W filter \textit{HST} image to a ground based \textit{R}-band image, to achieve the desired resolution in the central parts of the optical image. Neither the re-scaling procedure nor the MGE parameterisation is described in the paper, rendering us unable to perform any substantive further analysis of this issue. However, it is worth nothing that the overly-large $M/L_R$ may be contributing to the smaller SMBH mass of \citet{Rusli2013}, given the conservation of (total) dynamical mass during modelling.

The Schwarzschild model of \citet{Rusli2013} is a far more computationally-expensive and involved method than JAM, so one would ostensibly expect it to yield more realistic results. In addition, \citet{Rusli2013} considered models with and without a contribution from dark matter which, as mentioned in Section~\ref{subsection:uncertainties}, can be a source of uncertainty especially affecting the stellar mass component of models. However, the NGC~4751 SMBH mass and $M/L_R$ of \citet{Rusli2013} appear to be almost entirely unaffected by the addition of dark matter. In fact, the angular resolution of the SINFONI data (i.e.\ the FWHM of the PSF) of NGC~4751 is $0.22$~arcsec, whereas the diameter of the SoI (assuming the SMBH mass of \citealt{Rusli2013}) is $0.70$~arcsec, suggesting that the SINFONI data have an angular resolution sufficient to resolve the SoI. We would then expect the SMBH mass to be robust and less susceptible to any bias. If it is indeed the case that \citeauthor{Rusli2013}'s (\citeyear{Rusli2013}) SMBH mass is incorrect, as our JAM and molecular gas modelling results seem to suggest, it nevertheless highlights the aforementioned importance of cross-checks.


\subsection{The $M_{\text{BH}}$ -- $\sigma_\text{e}$ relation: comparison to the literature}

As outlined in Section~\ref{section:Introduction}, SMBH masses are often considered within the context of host galaxy -- SMBH mass correlations, that aid our understanding of their (co-)evolution. We thus compare our NGC~4751 SMBH mass measurements to the $M_{\text{BH}}$ -- $\sigma_\text{e}$ relation of \cite{Bosch_2016}. Despite being $2.0$ -- $2.5$ times larger than the mass inferred by \citet{Rusli2013}, both of our SMBH mass measurements inferred using the molecular gas method are still slightly under-massive given NGC~4751's stellar velocity dispersion, although both are also well within the scatter of the relation. Likewise, our stellar kinematic SMBH masses are still slightly under-massive but within the scatter of the $M_{\text{BH}}$ -- $\sigma_\text{e}$ relation.

The discrepancy between our SMBH mass measurements and those of \citet{Rusli2013} also highlight a long-standing issue in the study of SMBHs and host galaxy -- SMBH mass correlations. There are differences between SMBH masses determined using different kinematic tracers, but also, as can be seen in NGC~4751, between SMBH masses determined using the same tracer but different modelling methods/assumptions. It is commonly stated that ideally all these methods would be cross-checked against maser measurements, generally considered the gold standard of SMBH mass determination, to determine the accuracy and potential biases of the methods. However this gold standard status is more the exception than the rule for maser discs. Firstly, the rarity of masers and their strong bias towards relatively low-mass active galaxies (with correspondingly light SMBHs) makes the pursuit of such comparisons impossible. Secondly, maser discs often have a complicated velocity structure, less extension towards the SMBH, and generally few data points sampling the rotation curves. In fact, given this and the scarcity of targets, ALMA CO data measurements are arguably the most promising candidates for a new gold standard, extending to a broader set of targets while better sampling the rotation curves within the SMBH SoI \citep{Zhang_2024}.

\begin{figure}
    \centering
    \includegraphics[width=\columnwidth, trim={0.5cm 0 0 0}]{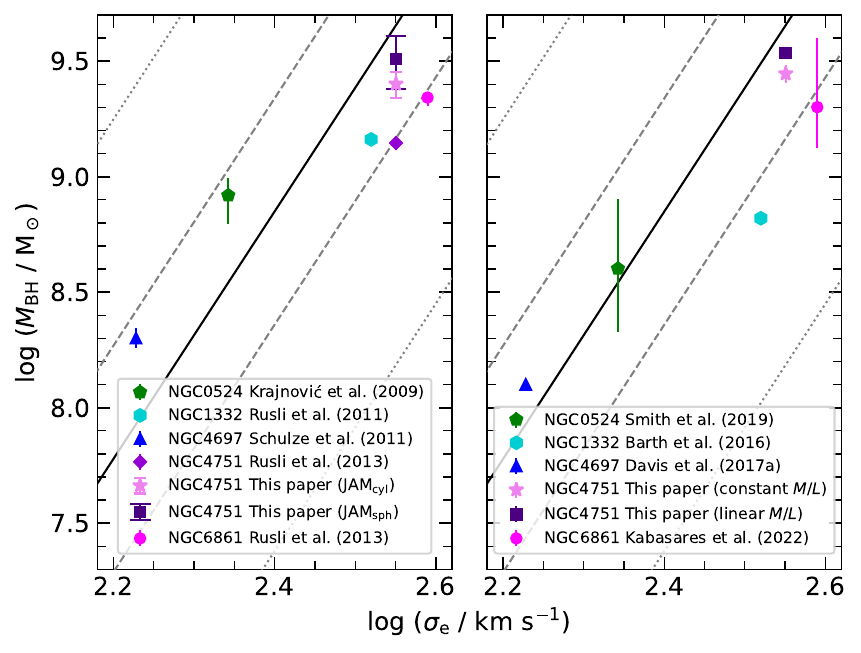}
    \caption{Best-fitting SMBH mass -- stellar velocity dispersion relation (black line) of \protect\citet{Bosch_2016}, with SMBH masses derived using both stellar kinematics (left) and molecular gas kinematics (right). Dashed and dotted grey lines show the $1\sigma$ and $3\sigma$ scatter of the relation, respectively. Error bars indicate $3\sigma$ confidence level uncertainties for \protect\cite{Krajnovic2009} and \protect\cite{WISDOM_IV}, $1\sigma$ uncertainties otherwise.
    }
    \label{fig:m-sigma}
\end{figure}

Aside from the dwarf galaxy NGC~404, that has a SMBH mass of $<10^6$~M$_\odot$, there are currently only five galaxies with both a molecular gas and a stellar kinematic SMBH mass measurement: NGC~524 \citep{Krajnovic2009, WISDOM_IV}, NGC~1332 \citep{Rusli_2010, Barth2016}, NGC~4697 \citep{Schulze_2011, Davis2017}, NGC~6861 \citep{Rusli2013, Kabasares_2022} and now NGC~4751 (\citealt{Rusli2013} and this paper). Figure~\ref{fig:m-sigma} compares the stellar (left panel) and molecular gas (right panel) kinematic measurements of the SMBH masses of these five objects. Whilst on average the molecular mass measurements tend to lie closer to the $M_{\text{BH}}$ -- $\sigma_\text{e}$ relation, four of the five galaxies have SMBH masses that are entirely consistent between the two methods (NGC~524, NGC~4751, NGC~4697 and NGC~6861). In addition, closer inspection of these results provides context and demands caution when making such comparisons. Indeed, the masses were determined using a variety of methods, with a variety of assumptions and a variety of data quality.

\citet{Krajnovic2009} used Schwarzschild modelling with the assumption of axisymmetry to model NGC~524. The data have a PSF FWHM of $0.23$~arcsec, which is a few times smaller than the SMBH $R_\text{SoI}$ of $0.6$~arcsec (here and throughout we will use the SMBH mass of the respective work when evaluating the SoI). However, as NGC~524 is face-on, it is challenging to estimate its inclination, as that involves additional assumptions about its intrinsic shape. Nevertheless, the assumed inclination of $20\degree$ is similar to that ($i=20\fdg6$) derived by \citet{WISDOM_IV}. Whilst the angular resolution of the \citet{WISDOM_IV} data ($R_\text{beam}=0.3$~arcsec) is comparable to their estimated SoI ($R_\text{SoI}=0.31$~arcsec), the molecular gas data of NGC~524 exhibit a large central hole, determined to have a radius $R_\text{hole}=0.52$~arcsec which, coupled with the low inclination angle, is likely responsible for the unusually large confidence intervals of the reported SMBH mass. Nevertheless, the SMBH masses derived using the two methods are consistent with each other within $3\sigma$.

NGC~1332 is one of the galaxies for which multiple results disagree with each other. This is particularly surprising as \citet{Barth2016} used the same stellar luminosity profile as \citet{Rusli_2010}. \citet{Rusli_2010} used Schwarzschild modelling with the assumption of axisymmetry, but also assumed an inclination $i=90\degree$. The PSF FWHM of the data is $0.14$~arcsec whilst  $R_\text{SoI}=0.38$~arcsec, so that sufficient resolution of the SoI was achieved. The molecular gas data of \citet{Barth2016} have a geometric mean resolution of $0.044$~arcsec, which again more-than-adequately resolves the $R_\text{SoI}$ of $0.23$~arcsec. The data show that NGC~1332 is almost but not exactly edge-on, with an inclination $i=84\fdg1$, so that the $i=90\degree$ assumption of \citet{Rusli2013} was inappropriate. However, it is unlikely that such a small change in inclination can lead to a factor of $\approx2$ shift in SMBH mass. We therefore conclude that the SMBH mass discrepancy does not have an easily identifiable dynamical cause.

\citet{Schulze_2011} also modelled NGC~4697 using the Schwarzschild method, assuming axisymmetry and an inclination $i=90\degree$. The $R_\text{SoI}$ of $0.45$~arcsec is fully resolved by the data, whose PSF FWHM is $0.08$~arcsec. The molecular gas data of \citet{Davis2017} have a synthesised beam of $0.53$~arcsec, which does not formally resolve the $R_\text{SoI}$ of $0.34$~arcsec. The dynamically-determined inclination of $76\fdg1$ suggests that \citeauthor{Schulze_2011}'s (\citeyear{Schulze_2011}) assumption of $i=90\degree$ is severely off. Whilst the results are formally in disagreement with each other, they were also determined assuming different distances. When the \citet{Schulze_2011} result is re-scaled to the distance assumed by \citet{Davis2017}, the resultant SMBH mass is $1.6\times10^8$~M$_\odot$, significantly closer to and consistent with that of \citet{Davis2017} within $3\sigma$ (despite the erroneous inclination assumption).

Lastly, \citet{Rusli2013} used the Schwarzschild method to model NGC~6861, also assuming axisymmetry and an inclination $i=90\degree$. The data have a PSF FWHM of $0.38$~arcsec, which fully resolves the $R_\text{SoI}$ of $0.38$~arcsec. The molecular gas modelling of \citet{Kabasares_2022} yields an inclination $i=73\degree$, once again suggesting that the convenient $i=90\degree$ assumption in Schwarzschild modelling is incorrect. Despite this, the SMBH masses derived through the two methods are entirely consistent with each other.

\begin{figure}
    \centering
    \includegraphics[scale = 0.68, trim = {1cm 0 0 0}]{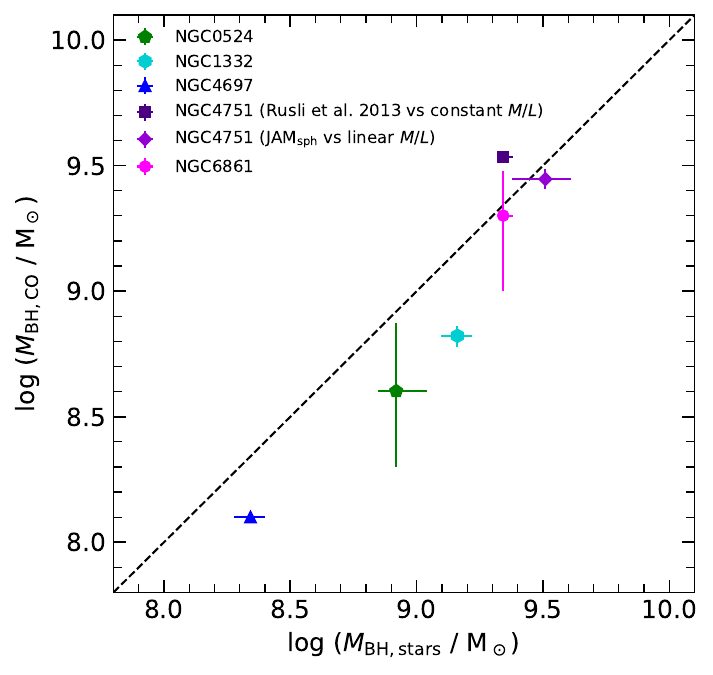}
    \caption{Cross comparison of SMBH masses derived using stellar kinematics and molecular gas. For clarity we have only included two data points for NGC~4751, representing the two extremes. One data point compares the least massive stellar kinematic mass estimate \citep{Rusli2013} and our most massive molecular gas mass estimate (our model with a constant $M/L$); the other compares the most massive stellar kinematic mass estimate (our model assuming JAM$_\text{sph}$) and our least massive molecular gas mass estimate (our model with a linearly-varying $M/L$). All data points comparing any other two methods would lie between those two. Error bars are as in Fig.~\ref{fig:m-sigma}.}
    \label{fig:mco vs mstar}
\end{figure}

Despite no clear trend with respect to the $M_{\text{BH}}$ -- $\sigma_\text{e}$ relation, Fig.~\ref{fig:mco vs mstar} suggests that molecular gas kinematics yields systematically lower SMBH mass estimates than stellar kinematics. It is hard to establish whether stellar kinematic modelling overestimates the masses or molecular gas modelling underestimates the masses, but molecular gas modelling is generally considered to be a lot simpler. The main assumption underlying molecular gas modelling is the presence of circular motions, which can be easily verified from the data themselves. Subsequently, the modelling is relatively straightforward and has considerable flexibility for implementing position angle and/or inclination warps. Both JAM and Schwarzschild modelling cannot rely on the simple assumption that the tracer population lies in a simple thin disc and they have to fit for the velocity anisotropy of the stars. However, the stars, unlike the gas, have the advantage of being unaffected by non-gravitational forces such as inflows, outflows and shocks. Both molecular gas and stellar methods can easily model radial $M/L$ gradients. Overall, more cross checks are needed to reach a conclusion on the reliability and possible biases of different SMBH mass determination methods.

\section{Conclusions}
\label{section:Conclusion}

High angular resolution ALMA observations were obtained and used to create a $^{12}$CO(3–2) data cube of the ETG galaxy NGC~4751. We presented dynamical models with two different $M/L$ radial profiles: constant and linearly varying. We estimated the stellar mass distribution using a MGE model of a \textit{HST} image and each of the $M/L$ profiles, and then forward modelled the molecular gas distribution and kinematics using \textsc{KinMS} and an MCMC framework. NGC~4751 has a regularly-rotating molecular gas disc with an inferred SMBH mass $M_\text{BH}=3.43^{+0.45}_{-0.44}[\text{stat},\,3\sigma]^{+0.22}_{-0.64}[\text{sys}]\times10^9$~M$_\odot$ and a stellar F160W filter $M/L$ $M/L_{F160W}=2.68\pm0.11[\text{stat},\,3\sigma]^{+0.10}_{-0.80}[\text{sys}]$~M$_\odot$/L$_{\odot,\text{F160W}}$ when assuming a constant $M/L$, and a SMBH mass $M_\text{BH}=2.79^{+0.75}_{-0.57}[\text{stat},\,3\sigma]^{+0.75}_{-0.45}[\text{syst}]\times10^9$~M$_\odot$ and a stellar F160W filter $M/L$ $\left(M/L_\text{F160W}\right)/\left(\text{M}_\odot/\text{L}_{\odot,\text{F160W}}\right)=3.07^{+0.27}_{-0.35}[\text{stat},\,3\sigma]^{+0.08}_{-1.14}[\text{sys}] -0.09^{+0.08}_{-0.06}[\text{stat},\,3\sigma]^{+0.08}_{-0.01}[\text{sys}]\,\left(R/\text{arcsec}\right)$ when assuming a radially linearly-varying $M/L$ (all statistical uncertainties at the $3\sigma$ confidence levels).

We additionally presented stellar kinematic SMBH mass estimates using the JAM method and SINFONI stellar kinematics. Assuming JAM$_\text{cyl}$ we obtained $M_\text{BH} = (2.52\pm 0.36)\times10^9$~M$_\odot$, while assuming JAM$_\text{sph}$ we obtained $M_\text{BH}=(3.24\pm0.87)\times10^9$~M$_\odot$.
The SMBH mass of the molecular gas model with a constant $M/L$ is statistically consistent with that of the model with a linearly-varying $M/L$ and JAM$_{\rm sph}$, whereas the SMBH mass of the molecular gas model with a linearly-varying $M/L$ is consistent with both JAM$_{\rm cyl}$ and JAM$_{\rm sph}$. The SMBH masses of the molecular gas model with a constant $M/L$ and JAM$_{\rm cyl}$ are marginally inconsistent with each other. 
All our measurements are larger than the previous stellar kinematic measurements of \citet{Rusli2013}, obtained using \citeauthor{Schwarzschild1979}'s (\citeyear{Schwarzschild1979}) orbit-superposition method. All of our SMBH masses are consistent with that predicted by the $M_{\text{BH}}$ -- $\sigma_\text{e}$ relation of \citet{Bosch_2016}. We conclude that the best-fitting SMBH mass of \citet{Rusli2013} is strongly excluded by our observations.


\section*{Acknowledgements}

PD acknowledges support from a Science and Technology Facilities Council (STFC) DPhil studentship under grant ST/S505638/1. MB was supported by STFC consolidated grant “Astrophysics at Oxford” ST/K00106X/1 and ST/W000903/1. TAD and IR acknowledge support from STFC grant ST/S00033X/1. This paper makes use of the following ALMA data: ADS/JAO.ALMA 2016.1.01135.S. ALMA is a partnership of ESO (representing its member states), NSF (USA), and NINS (Japan), together with NRC (Canada), MOST and ASIAA (Taiwan), and KASI (Republic of Korea), in cooperation with the Republic of Chile. The Joint ALMA Observatory is operated by ESO, AUI/NRAO, and NAOJ. This research made use of the NASA/IPAC Extragalactic Database (NED), which is operated by the Jet Propulsion Laboratory, California Institute of Technology, under contract with the National Aeronautics and Space Administration.


\section*{Data Availability}
 
The observations underlying this article are available in the ALMA archive, at \url{https://almascience.eso.org/asax/}, and in the Hubble Science Archive, at \url{https://hst.esac.esa.int/ehst/}.


\bibliographystyle{mnras}
\bibliography{BIB} 


\appendix

\section{Moment Maps}
\label{section:Appendix}

Comparison of the moment maps of the observed data and the best-fitting constant $M/L$ model of NGC~4751.


\begin{figure*}
    \centering
    \includegraphics[scale=0.45,trim={0.3cm 0 0 0}]{Figures/ngc4751_moment0.pdf}
    \hspace{0.5cm}
    \includegraphics[scale=0.45, trim={0 0 1cm 0}]{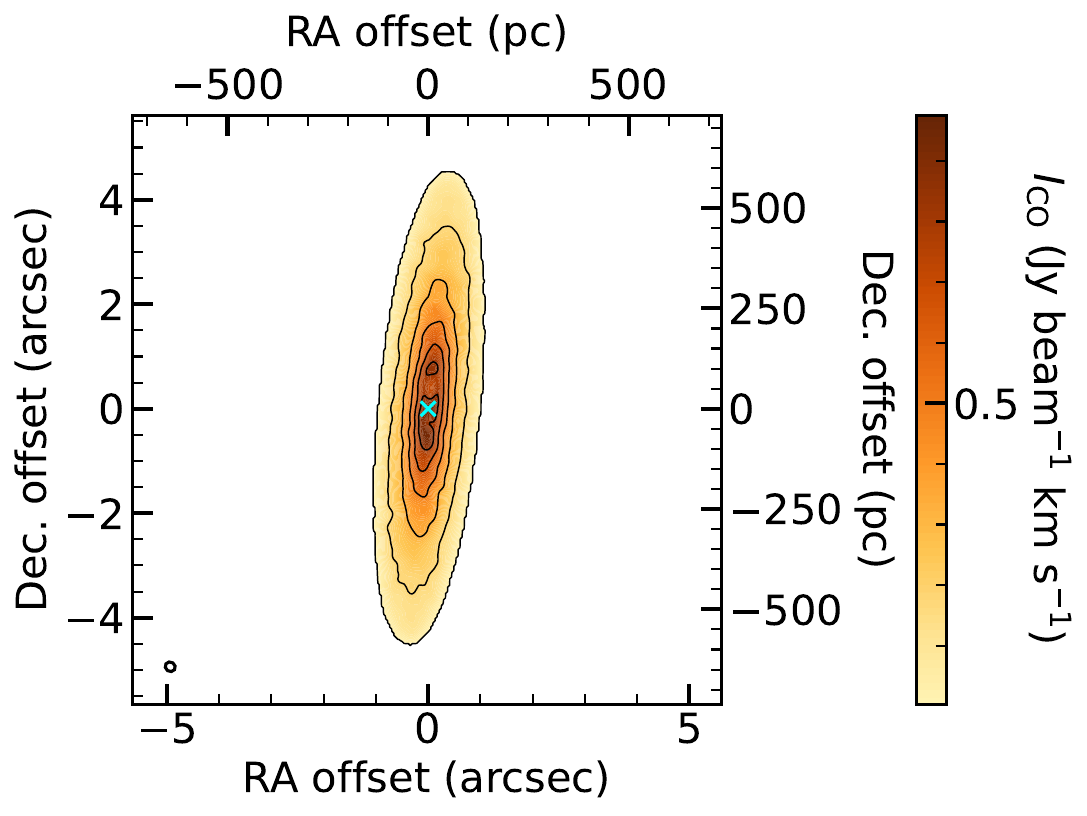}
    \newline
    \includegraphics[scale=0.45,trim={0.0cm 0 0.1cm 0}]{Figures/ngc4751_moment1.pdf}
    \hspace{0.3cm}
    \includegraphics[scale=0.45, trim={0.1cm 0 1.1cm 0}]{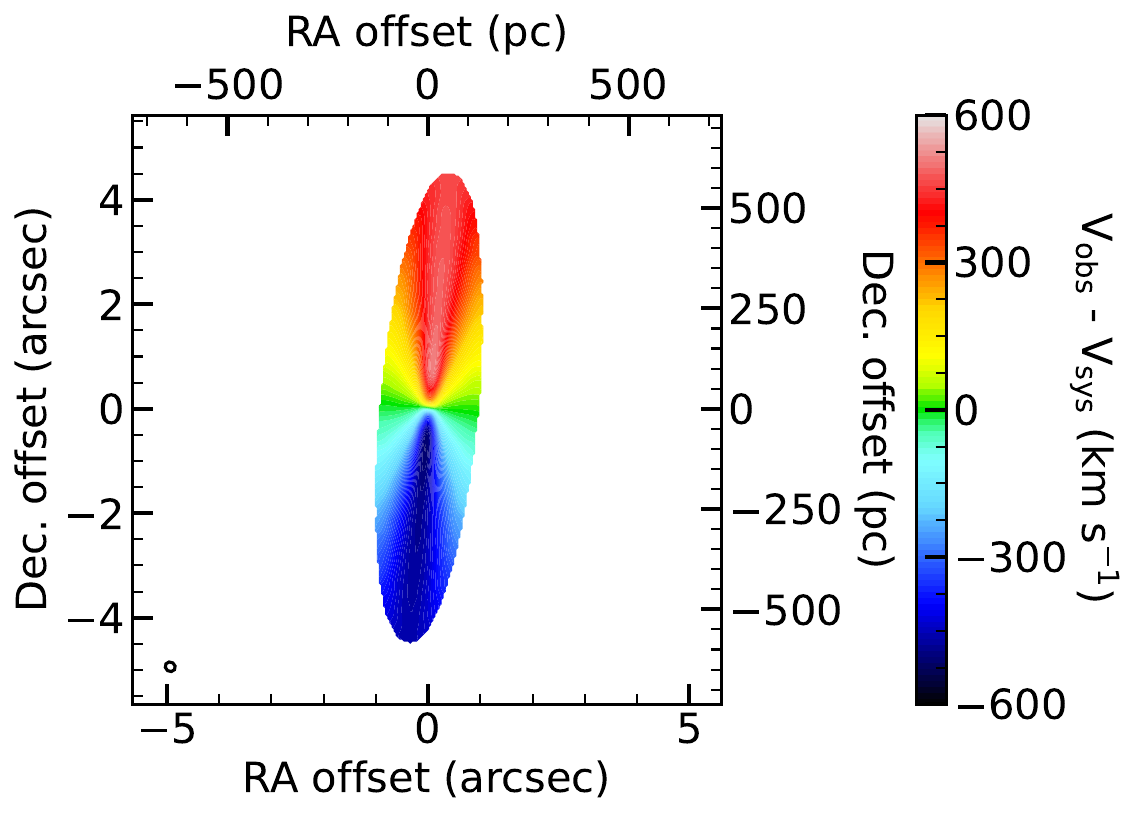}
    \newline
    \hspace*{-0.5cm}
    \includegraphics[scale=0.45,trim={0.1cm 0 0 0}]{Figures/ngc4751_moment2.pdf}
    \hspace{0.6cm}
    \includegraphics[scale=0.45, trim={0 0 1cm 0}]{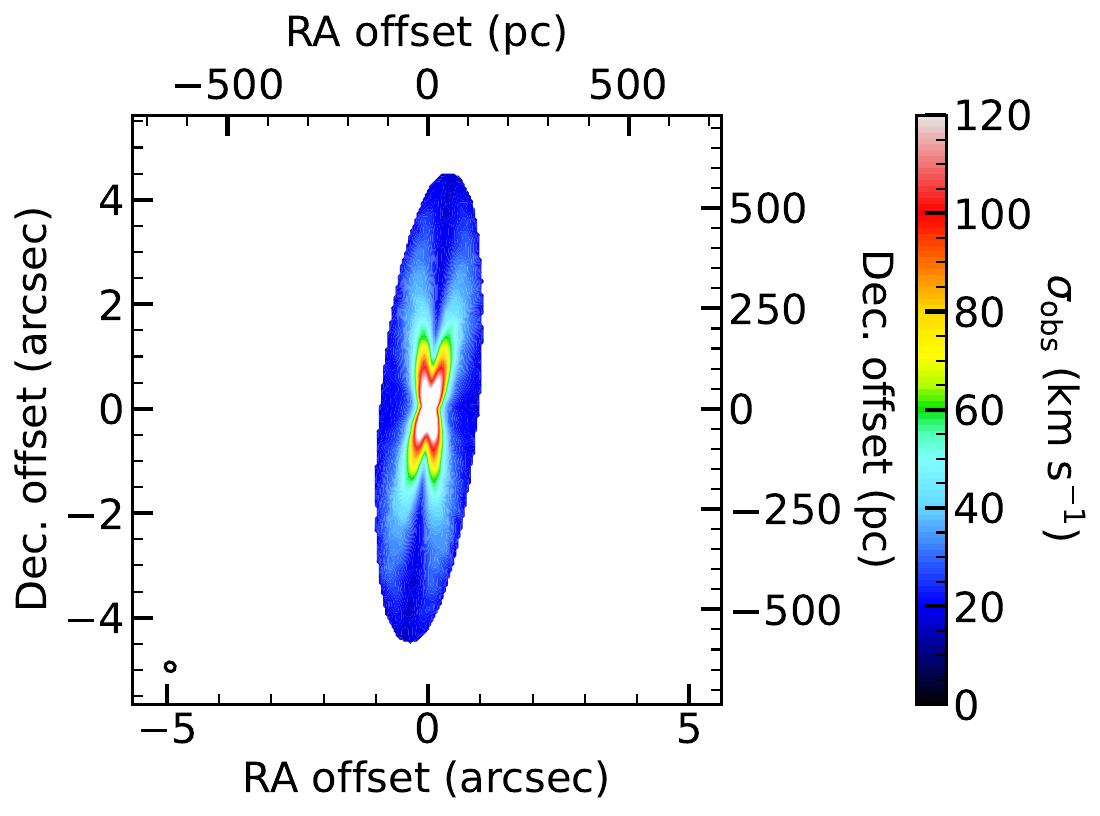}
    \caption{$^{12}$CO(3–2) zeroth-moment (integrated-intensity), first-moment (intensity-weighted mean line-of-sight velocity) and second-moment (intensity-weighted line-of-sight velocity dispersion) maps of NGC~4751. Moments extracted from the observed data cube are in the left panels, whilst those extracted from the simulated best-fitting data cube are in the right panels. The best-fitting data cube was created using the parameters of the exponential disc model with a constant $M/L$.}
    \label{fig:modelmoments}
\end{figure*}


\bsp	
\label{lastpage}


\end{document}